\documentclass[12pt, draftclsnofoot, onecolumn]{IEEEtran}
\usepackage[utf8]{inputenc}
\usepackage[cmex10]{amsmath}
\usepackage{upgreek}
\usepackage{booktabs}
\usepackage{epsfig}
\usepackage{latexsym}
\usepackage{multirow}
\usepackage{stfloats}
\usepackage{epstopdf}
\usepackage{color}  
\usepackage{tabularx} 
\usepackage{algorithm}
\usepackage{amssymb}
\usepackage{enumerate}
\usepackage{array}
\graphicspath{{./Figures/}}
\usepackage{color}
\usepackage{bbm}
\usepackage{bm}
\usepackage{cite}
\usepackage[tight,footnotesize]{subfigure}
\usepackage{balance}
\usepackage{mathrsfs}
\usepackage{verbatim}
\usepackage{dsfont}
\usepackage{verbatim}
\usepackage{tikz}
\usepackage{setspace}
\usepackage{diagbox}
\usepackage{multicol}
\usepackage{environ}
\usepackage{tikz}
\usepackage{amsmath}
\usepackage{stfloats}
\usepackage{algorithm}
\usepackage{algpseudocode}
\usepackage{amsmath}
\usepackage{graphics}
\usepackage{epsfig}
\usepackage{caption}
\allowdisplaybreaks[4]
\usepackage{amsmath}
\usepackage{amsthm}
\usepackage{authblk}

\newcommand\ffrac[2]{\frac{\displaystyle#1}{\displaystyle#2}}
\def\BibTeX{{\rm B\kern-.05em{\sc i\kern-.025em b}\kern-.08em
		T\kern-.1667em\lower.7ex\hbox{E}\kern-.125emX}}
\begin{document}
	\title{On Multiple-Antenna Techniques for Physical-Layer Range Security in the Terahertz Band}
	\author{Weijun~Gao,~\IEEEmembership{Graduate Student Member,~IEEE},
		Xuyang~Lu,~\IEEEmembership{Member,~IEEE},
		Chong~Han,~\IEEEmembership{Member,~IEEE}, and~Zhi~Chen,~\IEEEmembership{Senior Member,~IEEE}
		
		\thanks{
			\par
			Weijun Gao and Chong Han are with the Terahertz Wireless Communications (TWC) Laboratory, Shanghai Jiao Tong University, China (email:\{gaoweijun, chong.han\}@sjtu.edu.cn).
			\par 
			Xuyang Lu is with the Laboratory of Ultrafast Integrated Systems (LUIS), Shanghai Jiao Tong University, China (email: 	xuyang.lu@sjtu.edu.cn).
			\par Zhi Chen is with University of Electronic Science and Technology of China, China (email: chenzhi@uestc.edu.cn).
		}
	}
	{}
	\maketitle
	\thispagestyle{empty}
	\begin{abstract}
		Terahertz (THz) communications have naturally promising physical layer security (PLS) performance in the angular domain due to the high directivity feature brought by the ultra-massive multiple-antenna techniques.	
		However, traditional multiple-antenna techniques fail to combat eavesdroppers residing in the THz beam sector, even when the communication distances of legitimate users and eavesdroppers are different. This THz range security challenge motivates us to study new multiple-antenna techniques to provide THz range and angular security. In this paper, we first conduct a theoretical analysis of the secrecy capacity of the multiple antenna channel under the range security scenario. 
		Based on this, the frequency diverse array, as a candidate multiple-antenna technique, is proven ineffective in addressing the range security problem. 
		Then, motivated by the theoretical analysis, a novel widely-spaced array and beamforming design for THz range security are proposed, which realize communications in the near-field regions. A non-constrained optimum approaching (NCOA) algorithm is developed to
		achieve the optimal secrecy rate. Simulation results illustrate that under the range security scenario where the eavesdropper is inside the beam sector, our proposed widely-spaced antenna communication scheme can ensure a 6~bps/Hz secrecy rate when the transmit power is 10~dBm and the propagation distance is 10~m.
	\end{abstract}
	
	\section{Introduction}
	\subsection{Background}
	With an increasing demand for high-speed, reliable, and private information exchange, a new spectrum and advanced security technologies are awaited to be developed for next-generation wireless networks. 
	The Terahertz (THz) band, with frequencies ranging from 0.1 to 10~THz, is envisioned to realize multi-Gbps or even Tbps data rates owing to the abundant bandwidth resource~\cite{giordani2020toward,rappaport2019wireless,9766110}.
	In addition to the bandwidth merit, the potential of THz communications to improve information security is envisioned in~\cite{AKYILDIZ201416, eavesdropping,han2022molecular}. As an essential technique in addition to the conventional encryption-based methods, physical layer security (PLS) ensures information privacy at the physical layer of the wireless networks by exploiting transceiver and channel properties, including beamforming, fading, noise, and interference~\cite{aghdam2018overview}. Compared with the encryption-based methods, the PLS has the advantage of owning reduced overhead for secret key distribution. Moreover, PLS can ensure information secrecy even with an eavesdropper having infinite computational power. 
	
\begin{figure}
		\centering
		\includegraphics[width=0.8\textwidth]{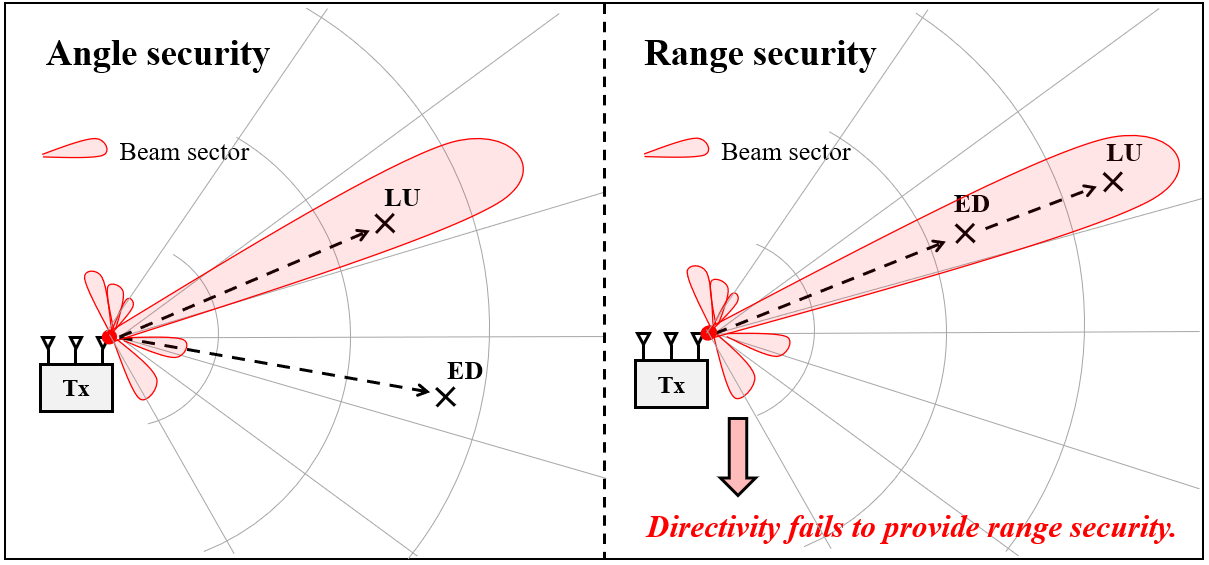}
		\caption{Illustration of range and angle security.}
		\label{fig:RA}
		\captionsetup{font={footnotesize}}
	\end{figure}

	With the equipment of ultra-massive multiple-input multiple-output (UM-MIMO) arrays for THz communications, the narrow and directional beams powered by multiple-antenna techniques bring substantial benefits to PLS~\cite{eavesdropping, myarticle1, myarticle2}. Typically, a THz beam is directed towards the legitimate user through the beam training and tracking processes.
	With an eavesdropper being located outside or inside the beam, the THz PLS can be divided into two categories, namely, \textit{angle security} and \textit{range security}, as illustrated in Fig.~\ref{fig:RA}. On the one hand, for the angle security, the aim is to prevent the eavesdropper from being located inside the beam sector, which improves the PLS in the angular domain. On the other hand, the range security aims at enhancing the secrecy rate when the legitimate user and the eavesdropper are both located inside the beam sector, i.e. when their propagation angles are the same while their ranges differ. Although the traditional multiple-antenna techniques of THz communications can enhance angle security, it fails to provide range security. Fortunately, there is a high probability that we only need to address the angle security problem since the THz beam width is narrow, e.g., approximately $1^\circ$ for a $32\times 32$ array,.
	
	Despite the promising confidentiality of THz angle security, solutions to the THz range security problem are still awaited to be explored for the following challenges~\cite{eavesdropping}. First, the signals received by the eavesdropper inside the beam benefit from the same antenna gain as legitimate users, which suggests that the THz directivity and traditional multiple-antenna techniques are ineffective in enhancing the range security. Second, due to the sparsity feature of the THz channel, the PLS technique using multi-paths to aid the security proposed in~\cite{petrov2019exploiting} might not be helpful. More critically, when Eve inside the beam sector is located nearer than the legitimate receiver, i.e., in close proximity, secure communication is significantly jeopardized since the secret codewords decodable by Bob are even easier decodable by the eavesdropper who has an even higher received signal-to-noise ratio (SNR)~\cite{myarticle2}.
	This research gap motivates this work aiming at addressing the range security for THz communications.  
	\subsection{Related Work}
	The fundamental limit of a wiretap channel model has been analyzed in~\cite{wyner1975wire} including a transmitter Alice (A), a legitimate receiver Bob (B), and an eavesdropper Eve (E). The key metric is characterized by the \textit{secrecy rate}, defined as the maximum data rate that can be transmitted reliably and confidentially.
	Traditional multiple-antenna technologies can only improve the secrecy rate in the angular domain. 
	For example, the secrecy capacity with the multiple antennas is calculated in~\cite{khisti2010secure}, while the multi-antenna-based secure beamforming techniques are proposed in~\cite{chen2016survey}, which controls the beamforming pattern to enlarge Bob's channel capacity while mitigating Eve's capacity. In~\cite{zheng2015multi,zheng2017physical,lv2018secure,ning2021joint}, as an extension to secure beamforming, the authors develop the artificial noise (AN)-aided beamforming technique to combat eavesdroppers in close proximity, where the AN signals are signals without carrying any information. In particular, AN signals are designed to be injected into the transmitted confidential signal to jam Eve's channel and enhance the secrecy rate. 
	
	Range security technologies, however, are rarely investigated for micro-wave band communications. Since the micro-wave channel contains rich scattering, angle security techniques can utilize the spatial degree of freedoms (SDoF) to mitigate eavesdropping attacks.
	However, the feature of limited SDoFs in THz communications makes the range security problem essential to be addressed. One useful range security technique in the literature is the receiver AN technique, where the receiver side sends the jamming signals while receiving the confidential signals~\cite{zheng2013improving, chen2018performance,yan2018secret}. On the downside, this technique requires a high-complexity self-interference cancellation (SIC) to transmit the AN signal and receive the information signal simultaneously. Another range security technique is the distance-adaptive molecular absorption modulation (DA-APM) scheme, which hides the signal under the molecular absorption peaks in the THz band~\cite{myarticle1}. This technique does not work when Eve is farther than Bob from Alice.
	
	In recent years, frequency diverse array (FDA) has been recognized as a good candidate to address the range security problem~\cite{lin2018physical}. 
	By introducing a small frequency offset among different antennas in the antenna array, the FDA shows a periodical range-, angle-, and time-varying beam pattern. FDA has been widely used in radar, navigation, and security applications by taking advantage of the time-varying property.
	For example, The FDA beamforming approach is proposed in\cite{lin2018physical} to safeguard wireless transmission for proximal Bob and Eve, where the FDA is used to distinguish two proximal receivers to increase the secrecy rate. 
	For the range security problem, FDA is recently envisioned as a potential technique to enhance information security thanks to its range-varying beam pattern. The effectiveness of the FDA-based techniques in providing THz range security requires investigation.
	
	\subsection{Motivations and Contributions}
	The goals of this paper are two-fold. First, we lay out a theoretical analysis of the multiple-antenna techniques under the range security condition and stress that multiple-antenna techniques operating in the far-field regions cannot enhance range security. 
	We revisit the FDA communication model as one particular design of multiple-antenna techniques and prove that previous FDA models in the literature~\cite{zheng2015multi,zheng2017physical,lv2018secure,ning2021joint} draw a wrong conclusion on its effectiveness on range security by missing the synchronization between array factor and symbol.   
	Second, motivated by the failure of multiple-antenna techniques operating in far-field regions, we propose a THz widely-spaced array (WSA) and a hybrid beamforming design as a near-field THz range security solution. Specifically, we present a multiple-antenna-assisted THz range security system model and develop the secrecy capacity, based on which the FDA system model is revisited and revised.
	Moreover, we propose a THz WSA and beamforming design to enhance the THz range security. An ultra-massive uniform planar array (UM-UPA) with widely-spaced antennas, instead of densely-packed antennas, is applied to provide THz range security.
	The trade-off analysis between the antenna size and the secrecy rate demonstrates that the THz WSA phased array can achieve an acceptable secrecy performance with a reasonable size thanks to the sub-millimeter wavelengths of THz waves.  
	The contributions of this paper are summarized as follows:
	\begin{itemize}
		\item \textbf{We perform theoretical analysis on multiple-antenna techniques for THz range security.} The secrecy capacity is derived based on which multiple-antenna techniques operating in the far-field regions are proven ineffective in providing range security. As a traditional candidacy for range security, the FDA communication model is analyzed and revised.
		\item \textbf{We propose a novel THz WSA and hybrid beamforming scheme to enhance the THz range security.} 
		A non-constrained optimum approaching (NCOA) algorithm is developed to solve the non-convex secrecy rate maximization problem, which determines the optimal hybrid beamformer design. Moreover, the trade-off between array size and secrecy rate enhancement of WSA is investigated. 
		
		\item \textbf{We evaluate our proposed scheme against the existing PLS techniques.}
		Monte-Carlo simulation demonstrates the convergence of the proposed NCOA algorithm.
		Extensive numerical results show that for a propagation distance at $10~\textrm{m}$ and transmit power of $10~\textrm{dBm}$, the proposed method can achieve a secrecy rate of $6~\textrm{bps/Hz}$, while the conventional antenna array fails to guarantee range security.
	\end{itemize}

	The remainder of the paper is organized as follows. Sec.~\ref{sec:sys} presents multiple-antenna-assisted THz secure communication system model. 
	Sec.~\ref{sec:fda} conducts a theoretical analysis on multiple-antenna techniques for THz range security, where the FDA technique is proven ineffective in providing range security. This conclusion is extended to the general multiple-antenna techniques operating in the far-field region. The proposed WSA communication and hybrid beamforming scheme are proposed in Sec.~\ref{sec:wsa}, where the NCOA hybrid beamforming algorithm is elaborated, and the trade-off between array size and maximum secrecy rate is investigated. Numerical results are described in Sec.~\ref{sec:num} to verify the convergence of the proposed algorithm and the secrecy rate enhancement compared with existing methods. The paper is concluded in Sec.~\ref{sec:concl}.
	
	\textit{Notations:}
	\rm matrices and vectors are denoted by boldface upper and lower case letters, respectively. $[\cdot]^+\triangleq \max(0,\cdot)$.
	$\langle\cdot, \cdot\rangle$ denotes inner product. $\ast$ is the convolution operation with respect to time $t$. $\mathbf{A}^T$, and $\mathbf{A}^\dagger$ are the transpose, and conjugate transpose of the matrix $\mathbf{A}$, respectively. 
	\begin{figure*}
		\centering
		\includegraphics[width=0.8\textwidth]{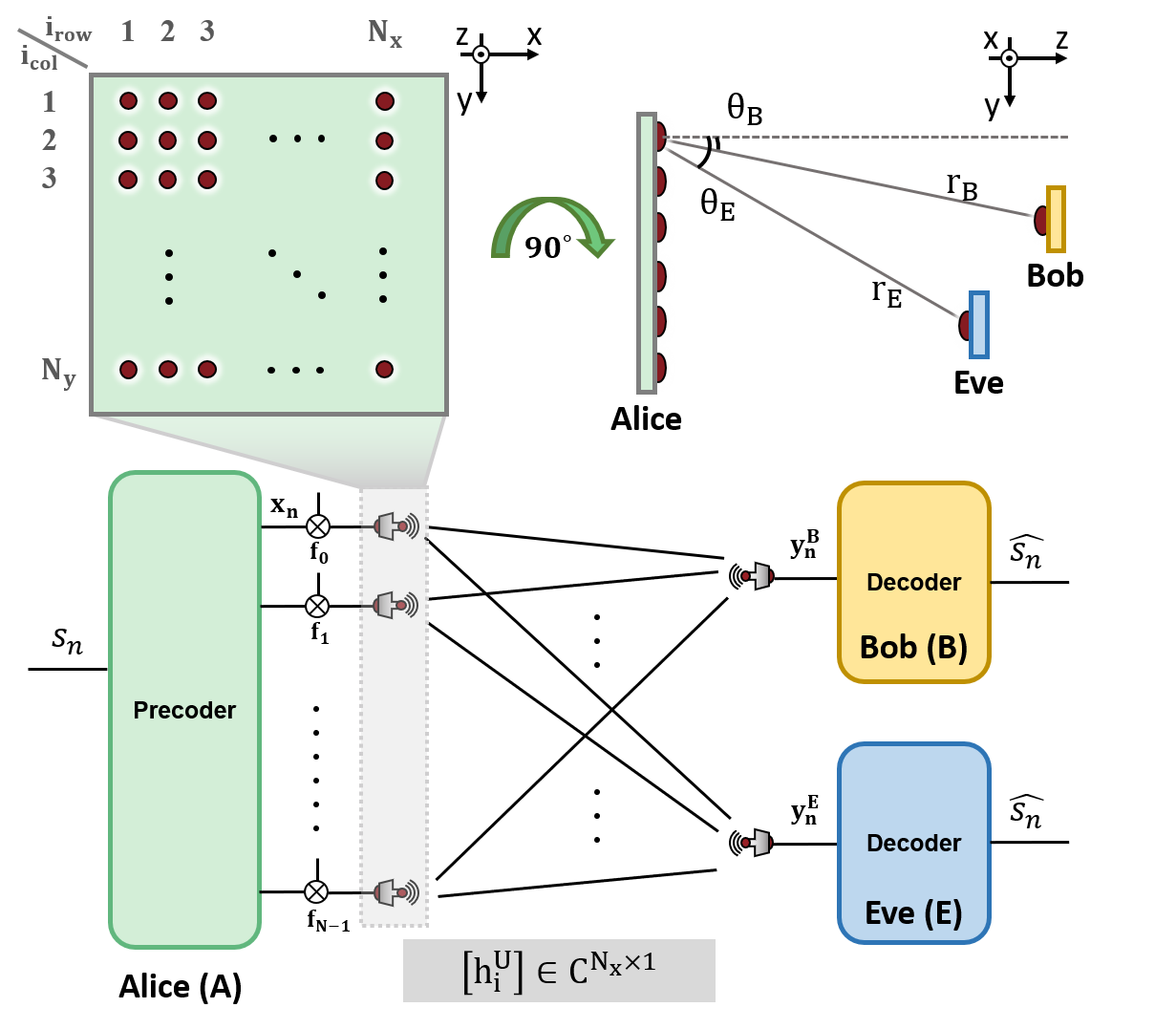}
		\caption{System model of multiple-antenna-assisted THz secure communication.}
		\label{fig:FDA_model}
	\end{figure*}  
	
	\section{System Model}~\label{sec:sys}
	In this section, we present a multiple-antenna-assisted THz secure communication model in Fig~\ref{fig:FDA_model}, which consists of a transmitter Alice (A), a legitimate receiver Bob (B), and an eavesdropper Eve (E). 
	Thanks to the small wavelength of the THz signal in the order of millimeter down to sub-millimeter, an ultra-massive uniform planar array (UM-UPA) is equipped at Alice to overcome the large propagation loss. Bob and Eve are assumed to be equipped with a single antenna. The UM-UPA is comprised of $N_{t}=N_x\times N_y$ antennas placed uniformly in a rectangular shape, with $N_x$ antennas in each row and $N_y$ antennas in each column. The antenna spacing between neighboring antennas is denoted by $d$.
	A three-dimensional Cartesian coordinate system is adopted, where the antenna in the $i_\textrm{row}^{\textrm{th}}$ row and the $i_\textrm{col}^{\textrm{th}}$ column is located at $((i_\textrm{row}-1)d,(i_\textrm{col}-1)d,0)$. The antenna index is represented by $i=N_x\times(i_\textrm{row}-1)+i_\textrm{col}$.
	The distance from the $i^{\textrm{th}}$ antenna at Alice to the receiver $U$ are denoted by $r_{\textrm{U},i}$.
	The receiving nodes including Bob and Eve, denoted by $\textrm{B}$ and $\textrm{E}$, are assumed to be located at $(r_{\textrm{U}}\sin\theta_{\textrm{U}}, 0,r_{\textrm{U}}\cos\theta_{\textrm{U}})$, where $\textrm{U}\in\{\textrm{B},\textrm{E}\}$, $r_{\textrm{U}}$ represents the propagation distance between Alice and $\mathrm{U}$, and  $\theta_\textrm{U}$ defines the propagation angle of the receiver $\textrm{U}$. 
	
	The normalized transmitted symbol stream is denoted by $s_n$ satisfying $\mathbb{E}[|s_n|^2]=1$, where $n$ represents the symbol index and $\mathbb{E}[\cdot]$ returns the expectation. The symbols are modulated into a baseband waveform ${m}\left(t\right)$ given by
	\begin{equation}
	{m}(t)=\sum_{n=-\infty}^{\infty} \sqrt{P}s_n\cdot g\left(t-nT_s\right),
	\label{eq:m2}
	\end{equation}
	where $P$ denotes transmit power, $t$ represents time, $g\left(t\right)$ represents a normalized pulse shape, and $T_s$ denotes a symbol time. We consider that the baseband bandwidth $B_g=1/T_s$ is smaller than the coherence bandwidth, which implies narrowband communications. For simplicity, we adopt the rectangular pulse waveform, which is given by
	\begin{equation}
	g(t)=g_{\textrm{rect}}(t)=
	\begin{cases}
	\frac{1}{\sqrt{T_s}}, &t\in [0,T_s),\\
	0, &\textrm{otherwise},
	\end{cases}
	\label{eq:rec_pul}
	\end{equation}
	By substituting~\eqref{eq:rec_pul} into~\eqref{eq:m2}, the baseband signal is expressed as
	\begin{equation}
	m(t)=\sqrt{\frac{P}{T_s}}s_{\lfloor t/T_s\rfloor}.
	\label{eq:rect}
	\end{equation}
	where $\lfloor x\rfloor$ returns the largest integer smaller or equal to $x$. The baseband waveform $m(t)$ is first precoded by a linear time-varying precoder $\mathbf{W}(t)\triangleq[w_1(t),\cdots,w_{N_t}(t)]$, satisfying $\sum_{i=1}^{N_t}|w_i(t)|^2=N_t$. The carrier frequency of the $i^\textrm{th}$ antenna is denoted by $f_i$, and the transmit signal represented by $\mathbf{x}=[x_1, \cdots, x_{N_{t}}]^T$ is therefore expressed as
	\begin{equation}
	\begin{aligned}
	x_{i}\left(t\right)&={m}\left(t\right)w_{i}\left(t\right)\textrm{e}^{-j2\pi f_{i} t},\\
	&=\sqrt{\frac{P}{T_s}}s_{\lfloor t/T_s\rfloor}w_{i}\left(t\right)\textrm{e}^{-j2\pi f_{i} t},
	\end{aligned}
	\label{eq:xi_fda}
	\end{equation}
	where $x_i(t)$ denotes the signal transmitted by the $i^{\textrm{th}}$ antenna of Alice.
	Due to the sparsity of the THz channel and the high directivity, the path gains of the reflected, scattered, and diffracted paths are negligibly weak compared to the line-of-sight (LoS) path. Thus, we consider to model the THz channel impulse response $h_{\textrm{U},i}\left(t\right)$ from the $i^\textrm{th}$ antenna of Alice to node $\textrm{U}\in\{\textrm{B}, \textrm{E}\}$as the result of the LoS path only~\cite{han2014multi, Peng2020channel, chen2021millidegree}, which is represented as
	\begin{equation}
	h_{\textrm{U},i}\left(t\right)=\alpha\left(f_i,r_{\textrm{U},i}\right)\delta\left(t-\frac{r_{\textrm{U},i}}{c}\right),
	\label{eq:hlos}
	\end{equation}
	where $\alpha\left(f_i,r_{\textrm{U},i}\right)=\frac{c}{4\pi f_ir_{\textrm{U},i}}$ denotes the free-space LoS path gain according to Friis' law. $c$ denotes the speed of light. The delay is $\frac{r_{\textrm{U},i}}{c}$ and  $\delta\left(\cdot\right)$ stands for the Dirac Delta function. To simplify the notations, we use $\alpha(r_{\textrm{U},i})$ to represent $\alpha(f_i,r_{\textrm{U},i})$.
	The superimposed received signal $y_{\textrm{U}}\left(t\right)$ at node $\textrm{U}$ from the transmit antenna array is thus given by
	\begin{equation}
	y_{\textrm{U}}\left(t\right)=\sum_{i=1}^{N_{t}}x_{i}\left(t\right)\ast h_{\textrm{U},i}\left(t\right) + n_{\textrm{U}}\left(t\right), 
	\label{eq:yr}
	\end{equation}
	where $n_{\textrm{U}}(t)$ stands for the additive Gaussian white noise (AWGN) noise variance $\sigma_{\textrm{U}}^2$. In this work, we assume that the noise variances for Bob and the Eve are the same and both equal to $\sigma^2$, i.e., $\sigma_{\textrm{B}}^2=\sigma_{\textrm{E}}^2=\sigma^2$.
	
	According to the relationship between the antenna array size and the communication distance, multiple-antenna communications can be classified into near-field and far-field communications. If the propagation distance of the receiving node $\textrm{U}$ is larger than the Fraunhofer distance~\cite{myers2021infocus, yuhang2021hybrid}, i.e., $d\ll
	8r_{\textrm{U}}^2/\lambda$, the node $\textrm{U}$ is considered to be in the far-field region of the transmitter.  
	This distance relationship determines how we could further simplify the expression of the received signal represented in~\eqref{eq:rect}-\eqref{eq:yr}. In general, we represent the distance $r_{\textrm{U},i}$ as 
	\begin{equation}
	r_{\textrm{U},i}=r_{\textrm{U}}+\mathcal{L}(i,\theta_{\textrm{U}},r_{\textrm{U}}),
	\label{eq:rui_init}
	\end{equation}
	where $\mathcal{L}(i,\theta_{\textrm{U}},r_{\textrm{U}})$ denotes the residual propagation distance of the $i^\textrm{th}$ antenna to the receiver. This residual term depends on the antenna index $i$, the propagation angle $\theta_{\textrm{U}}$, and the propagation distance $r_{\textrm{U}}$. If the receiver is located in the far-field region of the transmitter antenna array, the following far-field approximation is satisfied by neglecting the high-order term related to $r_\textrm{U}$ as
	\begin{equation}
	r_{\textrm{U},i}\approx r_{\textrm{U}}+\mathcal{L}_{\textrm{p}}(i,\theta_{\textrm{U}}),
	\label{eq:ruij}
	\end{equation}
	where 
	$\mathcal{L}_{\textrm{p}}(i,\theta_\textrm{U})\ll r_{\textrm{U}}$ is range-invariant and only depends on the antenna index $i$ and the angle $\theta_{\textrm{U}}$. For example, if the uniform space $d$ satisfies $d\le c/2f_i$, the residual term of a linear antenna array can be expressed as $\mathcal{L}_{\textrm{p}}(i,\theta_{\textrm{U}})=id\sin\theta_{\textrm{U}}$ and that of a planar antenna array can be represented by $\mathcal{L}_{\textrm{p}}(i,\theta_{\textrm{U}})=-(i_{\textrm{row}}-1)d\sin\theta_{\textrm{U}}$. The counterpart of the far-field communication is defined as near-field communication, where the approximation~\eqref{eq:ruij} does not apply. Since the THz antenna array is in the order of several millimeters, most THz communication systems are far-field communications.
	\section{Theoretical Analysis of Multiple-antenna Techniques for Terahertz Range Security}\label{sec:fda}
	In this section, we theoretically investigate how multiple-antenna techniques can address the THz range security problem. To reach this goal, we first analyze the secrecy capacity of the multiple-antenna secure channel under the THz range security scenario. This secrecy capacity analysis provides an upper-bound secrecy performance of general multiple-antenna techniques. Then, we demonstrate that the FDA technique, as an advocated multiple-antenna technique in the literature for range security, is ineffective in enhancing THz range security.
	\begin{figure*}
		\centering
		\includegraphics[width=0.9\textwidth]{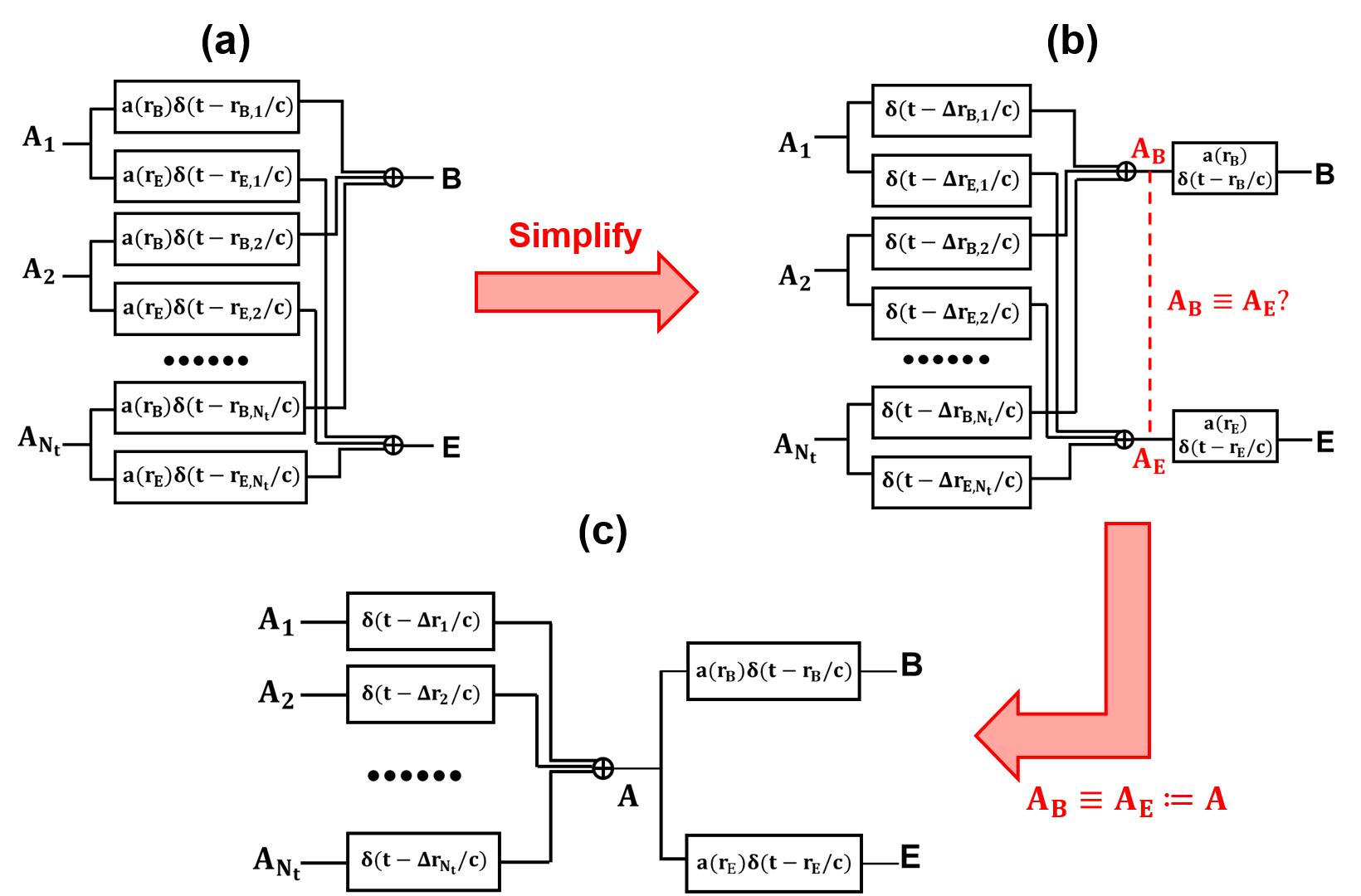}
		\caption{System block diagram transformation of the multiple-antenna secure communication system. (a) Original system block diagram. (b) First-transformed system block diagram. (c) Second-transformed system block diagram.}
		\captionsetup{font={footnotesize}}
		\label{fig:system_illu}
	\end{figure*} 
	\subsection{Secrecy Capacity Analysis Under Range Security Scenario}~\label{sec:SCA}
	We first theoretically derive the secrecy capacity of the multiple-antenna secure communication system under the range security scenario. Instead of assuming any specific precoding or up-converting method at the transmitter or receiver side, we focus on the original multiple-antenna secure channel. 
	The original channel block diagram is shown in (a) of Fig.~\ref{fig:system_illu}, where nodes $\{\textbf{A}_i\}, i\in[1,N_t]$ represent the antennas in the transmit antenna array. Node $\textbf{B}$ and $\textbf{E}$ represent the single-antenna Bob and Eve, respectively. The channel impulse response from $\textbf{A}_i$ to $\textbf{U}\in\{\textbf{B},\textbf{E}\}$ is expressed by $a(r_\textrm{U})\delta(t-\frac{r_{\textrm{U},i}}{c})$. 
	By substituting $r_{\textrm{U},i}=r_{\textrm{U}}+\Delta r_{\textrm{U},i}$, the system block diagram in Fig.~\ref{fig:system_illu}(a) can be simplified to the block diagram in Fig.~\ref{fig:system_illu}(b), due to the following relationship
	\begin{subequations}
		\begin{align}
		y_\textrm{U}(t)&=\sum_{i=1}^{N_t} x_i\left(t\right)\ast  a(r_{\textrm{U}})\delta\left(t-\frac{r_{\textrm{U},i}}{c}\right)\\
		&=\left(\sum_{i=1}^{N_t} x_i\left(t\right)\ast  \delta\left(t-\frac{\Delta r_{\textrm{U},i}}{c}\right) \right)\ast a(r_{\textrm{U}})\delta\left(t-\frac{ r_{\textrm{U}}}{c}\right), 
		\label{eq:1}
		\end{align}
	\end{subequations}
	where $x_i(t)$ represents the transmitted signal at node $\textbf{A}_i$, and $y_\textrm{U}(t)$ denotes the received signal at node $\textbf{U}$. Two intermediate nodes representing the superimposed signals at Bob and Eve in Fig.~\ref{fig:system_illu}(b) are defined as $\textbf{A}_{\textbf{B}}$ and $\textbf{A}_{\textbf{E}}$, respectively. As the basic assumption for the range security problem, the propagation angles of Bob and Eve are equal, i.e., $\theta_{\textrm{B}}=\theta_{\textrm{E}}$. Additionally, as described in Sec.\ref{sec:sys}, we assume that the receiving nodes $\mathbf{B}$ and $\mathbf{E}$ are in the far-field region of the antenna array. Therefore, $\Delta r_{\textrm{B},i}=\Delta r_{\textrm{E},i}$, and the signals at node $\textbf{A}_{\textbf{B}}$ and $\textbf{A}_{\textbf{E}}$ are equal. We can superimpose the two nodes and their corresponding  overlapped parts in (b), as depicted in Fig.~\ref{fig:system_illu}(c). Fig.~\ref{fig:system_illu}(c) shows a cascaded wiretap channel which exhibits as two Markov chains, e.g., 
	$\{\textbf{A}_i\}\to\textbf{A}\to \textbf{B}$ and
	$\{\textbf{A}_i\}\to\textbf{A}\to \textbf{E}$.
	
	Based on the system transformation from Fig.~\ref{fig:system_illu}(a) to Fig.~\ref{fig:system_illu}(c), the secrecy capacity $C_s$ of the multiple-antenna secure communication system can be derived according to the property of Markov chains, as
	\begin{equation}
	\begin{aligned}
	C_{s}&=\max_{[\textbf{A}_i]}I([\textbf{A}_i];\textbf{B})-\max_{[\textbf{A}_i]}I([\textbf{A}_i];\textbf{E})\\
	&\le \max_{\textbf{A}}I(\textbf{A};\textbf{B})-\max_{\textbf{A}}I(\textbf{A};\textbf{E})\\
	&=\log\left(1+\frac{P_{\textrm{A}}a(r_\textrm{B})^2}{\sigma_{\textrm{B}}^2}\right)-\log\left(1+\frac{P_{\textrm{A}}a(r_\textrm{E})^2}{\sigma_{\textrm{E}}^2}\right),
	\end{aligned}
	\label{eq:secrecy_rate_proof}
	\end{equation}
	where $P_{\textrm{A}}$ denotes the maximum signal power of the combined signals at node $\textbf{A}$. \eqref{eq:secrecy_rate_proof} implies that the secrecy rate from node $[\textbf{A}_i]$ is upper-bounded by the secrecy rate from node $\textbf{A}$ and
	\begin{equation}
	P_{\textrm{A}}=\max\mathbb{E}\left[\left|\sum_{i=1}^{N_t}x_i(t-\Delta r_i/c)\right|^2\right]\le \sum_{i=1}^{N_t}\max\mathbb{E}[|x_i(t)|^2]= N_{t}P,
	\end{equation}
	whose secrecy capacity is equal to that of a fully-digital beamforming scheme. This result proves that any signal processing and multiple antenna techniques operating in the far-field regions cannot provide THz range security since the traditional fully-digital beamforming technique can reach the secrecy capacity. 
	
	\begin{figure*}
		\centering
		\includegraphics[width=0.7\textwidth]{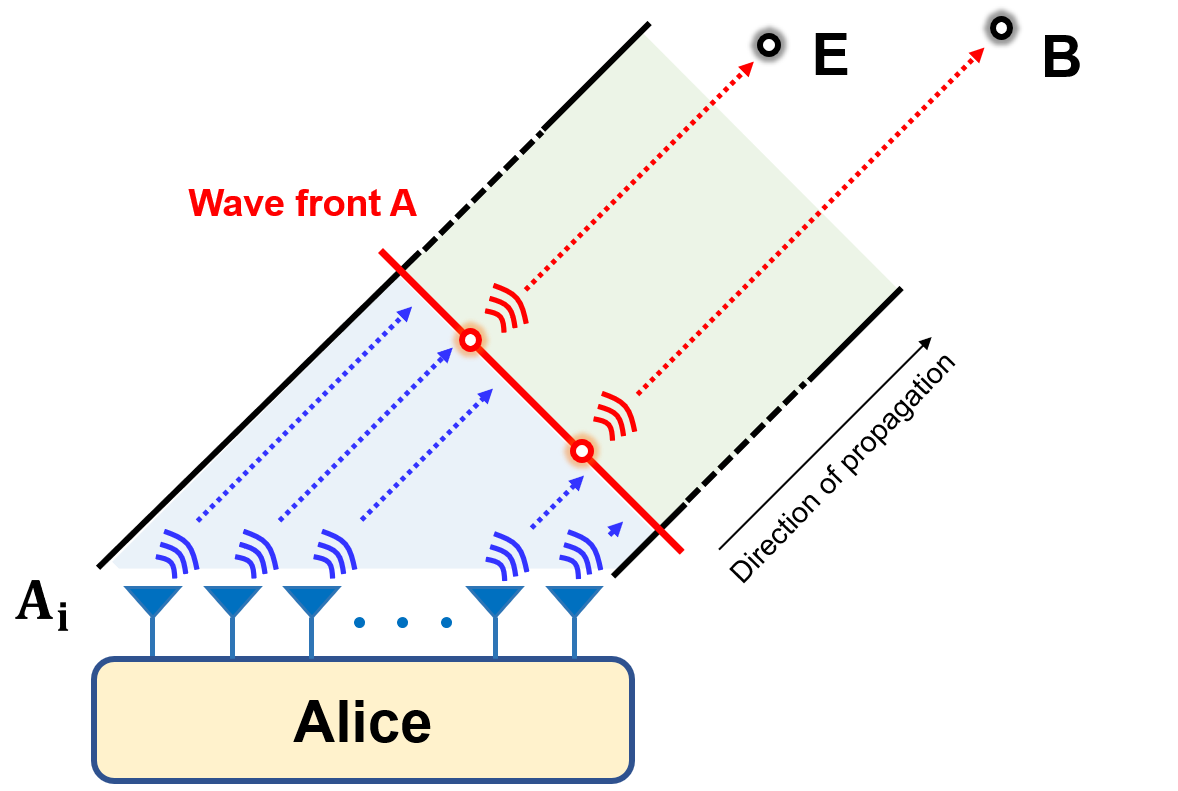}
		\caption{Physical meaning of the intermediate node $\textbf{A}$ in the Markov chain.}
		\label{fig:Nearfield}
	\end{figure*}  
	
	Further explanation from the electromagnetic wave propagation perspective is provided as follows.
	In Fig.~\ref{fig:system_illu}(c), the intermediate node $\textbf{A}$ can be interpreted as the wavefront of the transmitted beam. The THz LoS signal propagation from the source to far-field regions can be divided into two phases, as depicted in Fig.~\ref{fig:Nearfield}. In the first phase, the transmitted signal propagates from each antenna to the plane perpendicular to the direction of propagation in the near-field region, i.e., the wavefront $\mathbf{A}$. In the second phase, the combined signal further propagates and enters the far-field regions. According to the Markov chain theory, for all receiving nodes located in the far-field regions with the same propagation angle, the secrecy capacity of a MIMO channel is upper bounded by the SISO channel from the wavefront to the receiving node. This wavefront acts like a bottleneck of the wiretap channel, preventing multiple-antenna techniques from distinguishing Bob and Eve in the far-field region. Therefore, any multiple-antenna schemes operating in the far-field region cannot enhance THz range security.
	\subsection{Can Frequency Diverse Array Provide Terahertz Range Security?}\label{sec:fda_new}
	FDA introduces different small frequency offsets on each antenna to obtain a range-, angle-, and time-dependent array factor, which can steer the beam automatically~\cite{wang2015frequency}. The range-dependency of the beam can simultaneously enhance the received SNR at Bob while mitigating the received SNR at Eve since the antenna gain of FDA periodically changes with the distance~\cite{lin2018physical}. Based on this phenomenon, researchers explore FDA-based multiple-antenna techniques to achieve range security, and they assume that Bob and Eve are in the far-field regions of the FDA. 
	However, this idea contradicts our secrecy capacity analysis in Sec.~\ref{sec:SCA} since their derived data rate exceeds our derived secrecy capacity. Therefore, we rigorously revisit the traditional FDA range security model and revise the model to demonstrate that the FDA cannot provide THz range security.  
	\subsubsection{Traditional Frequency Diverse Array Range Security Model}
	In the traditional FDA model adopted in~\cite{cheng2019wfrft,cheng2019svd,qiu2019multi,ji2019physical,wang2021secrecy}, the carrier frequency of the $i^{\textrm{th}}$ antenna is assumed to be $f_i=f_0+\Delta f_i$, where $\Delta f_i\ll f_0$. By assuming the all-one matrix for the precoding matrix and far-field condition $\mathcal{L}(i,\theta_{\textrm{U}})=id\sin\theta_{\textrm{U}}$, the equivalent antenna gain, i.e., the array factor (AF), is given by~\cite{cheng2019wfrft}
	\begin{equation}
	AF(t,\theta_{\textrm{U}},r_{\textrm{U}})=\sum_{i=1}^{N_t}\exp\left\{j2\pi \left[\Delta f_i\left(t-\frac{r_{\textrm{U}}}{c}\right)+(f_0+\Delta f_i)\frac{id\sin\theta_{\textrm{U}}}{c}\right]\right\},
	\label{eq:fda}
	\end{equation}
	and the secrecy rate is given by 
	\begin{equation}
	R_s=\log\left(1+\frac{P|AF(t,\theta,r_{\textrm{B}})|^2|\alpha(r_{\textrm{B}})|^2}{\sigma^2}\right)-\log\left(1+\frac{P|AF(t,\theta,r_{\textrm{E}})|^2|\alpha(r_{\textrm{E}})|^2}{\sigma^2}\right),
	\label{eq:fda_rs}
	\end{equation}
	where $\theta_{\textrm{B}}=\theta_{\textrm{E}}=\theta$ for range security scenario.
	Since~\eqref{eq:fda} depends on range parameter $r$, it can be designed that the AF at Bob can be larger than AF at the Eve, which seems to be beneficial to range security.	
	
	However, the derivation from the array factor in~\eqref{eq:fda} to the secrecy rate in~\eqref{eq:fda_rs} adopts an inappropriate assumption.
	it is incorrectly assumed that the array factor $AF(t,\theta,r_{\textrm{B}})$ and $AF(t,\theta,r_{\textrm{E}})$ at the same time $t$ refer to the same symbol.
	Suppose Alice transmits a secret symbol at time $t_0$, then Bob and Eve can receive this symbol at time $t_0+\frac{r_{\textrm{B}}}{c}$ and $t_0+\frac{r_{\textrm{E}}}{c}$, respectively. Therefore, when we analyze the array factor at the receiver sides, $AF(t,\theta,r_\textrm{B})$ represents the AF of the symbol transmitted at $t-\frac{r_{\textrm{B}}}{c}$ while $AF(t,\theta,r_\textrm{B})$ stands for the AF of the symbol transmitted at $t-\frac{r_{\textrm{E}}}{c}$. These two AFs do not refer to the same symbol transmitted from Alice, and therefore it is incorrect to compute the secrecy capacity based on them.
	\begin{figure}
		\centering
		\includegraphics[width=4\textwidth/5]{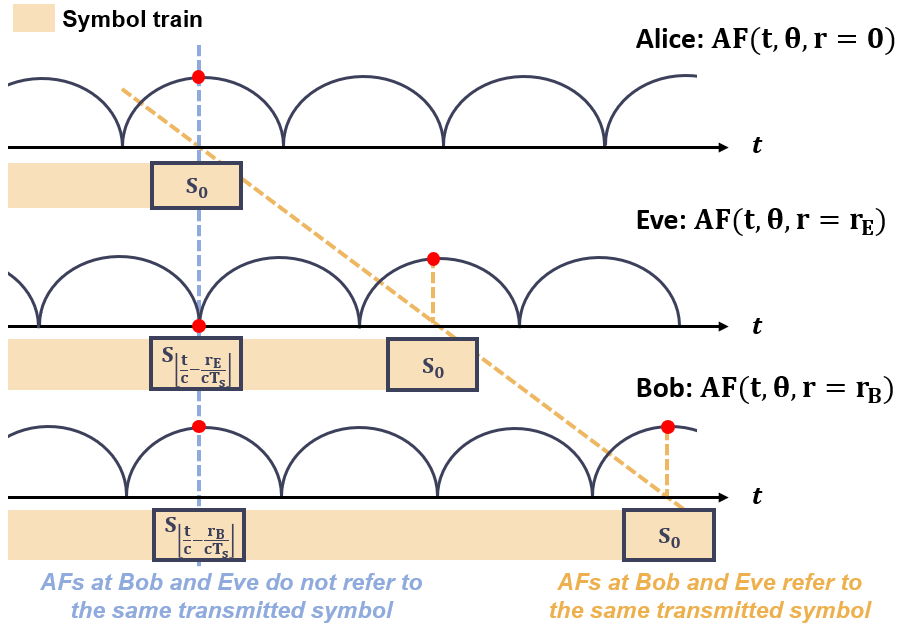}
		\caption{Illustration of the synchronization problem between AF and symbol in traditional FDA range security model.}
		\captionsetup{font={footnotesize}}
		\label{fig:symbol_AF}
	\end{figure}
	This example is illustrated in Fig.~\ref{fig:symbol_AF}. The blue line captures the derived traditional AF at Alice, Bob, and Eve, and we observe that it refers to different symbols. The correct way to study the secrecy rate is to track one secret symbol and focus on the AF of the same symbol received by Bob and Eve at different times, as the yellow line illustrates.
	Specifically in this example, we should compare the AF of the symbol at Bob and Eve at $t+\frac{r_{\textrm{B}}}{c}$ and $t+\frac{r_{\textrm{E}}}{c}$, respectively, which leads to the following observation that
	\begin{equation}
	AF\left(t_0+\frac{r_{\textrm{B}}}{c},\theta,r_{\textrm{B}}\right)=
	AF\left(t_0+\frac{r_{\textrm{E}}}{c},\theta,r_{\textrm{E}}\right).
	\end{equation}
	Therefore, the AF of one symbol created by the FDA is the same at Bob and Eve, which cannot provide range security.
	
	The illustration from the beam pattern point of view is provided as follows.
	Since the confidential message is modulated on the EM wave, which travels at the speed of light, we should investigate the joint range-time beam pattern of FDA instead of a ``snapshot" one. Unfortunately, although the FDA has a range-varying and time-varying beam pattern, the joint range-time beam pattern propagating at the speed of light is constant, as depicted in Eq.~(2) of~\cite{lin2018physical}, where the FDA beam pattern is only dependent on the term $t-\frac{r_{\textrm{U}}}{c}$. The interpretation of the FDA beam pattern in previous related studies neglects the synchronization between symbol and array factor, and therefore their secrecy rate analysis is problematic. 
	\subsubsection{Revised Frequency Diverse Array Range Security Model}
	We first rigorously derive the end-to-end received signals at Bob and Eve.
	The superimposed signal at the receiver $\textrm{U}$ can be computed by
	\begin{subequations}
		\begin{align}
		&\notag y_{\textrm{U}}(t,\theta,r_{\textrm{U}})\\
		&=\sqrt{\frac{P}{T_s}}\sum_{i=1}^{N_t} s_{\lfloor t/T_s\rfloor} w_i(t)\alpha(r_{\textrm{U},i})\exp\left\{-j2\pi f_{i} t\right\}*\delta\left(t-\frac{r_{\textrm{U},i}}{c}\right)
		\label{eq:yu1}
		\\
		&=\sqrt{\frac{P}{T_s}}\sum_{i=1}^{N_t}s_{\lfloor \frac{t}{T_s}-\frac{r_{\textrm{U},i}}{cT_s}\rfloor} w_i\left(t-\frac{r_{\textrm{U},i}}{c}\right)\alpha(r_{\textrm{U},i})\exp\left\{-j2\pi (f_0+\Delta f_i) \left(t-\frac{r_{\textrm{U},i}}{c}\right)\right\}
		\label{eq:yu2}
		\\
		&\approx\sqrt{\frac{P}{T_s}} s_{\lfloor \frac{t}{T_s}-\frac{r_{\textrm{U}}}{cT_s}\rfloor} \alpha(r_{\textrm{U}})\sum_{i=1}^{N_t}w_i\left(t-\frac{r_{\textrm{U}}}{c}\right)\exp\left\{-j2\pi (f_0+\Delta f_i) \left(t-\frac{r_{\textrm{U},i}}{c}\right)\right\}
		\label{eq:yu3}
		\\
		&\approx\sqrt{\frac{P}{T_s}} s_{\lfloor \frac{t}{T_s}-\frac{r_{\textrm{U}}}{cT_s}\rfloor} \alpha(r_{\textrm{U}})
		\textrm{e}^{-j 2\pi f_0\left(t-\frac{r_{\textrm{U}}}{c}\right)}
		\sum_{i=1}^{N_t}w_i\left(t-\frac{r_{\textrm{U}}}{c}\right)
		\exp\left\{-j2\pi \left[f_0\frac{\mathcal{L}_p(i,\theta)}{c}- \Delta f_i\left(t-\frac{r_{\textrm{U}}}{c}\right) \right] \right\}
		\label{eq:yu4}
		\end{align}
		\label{eq:FDA}
	\end{subequations}
	In~\eqref{eq:yu1}, $\alpha(r_{\textrm{U},i})$ represents the channel gain and the superimposed signal is expressed by the summation of signals transmitted from each antenna . In~\eqref{eq:yu2}, the convolution with the Dirac delta function $\delta\left(t-\frac{r_{\textrm{U},i}}{c}\right)$ can be simplified by replacing $t$ as $t-\frac{r_{\textrm{U},i}}{c}$.
	The approximation in~\eqref{eq:yu3} applies the assumption that the symbol, the precoding matrix, and the path loss are uniform for all the antenna pairs, i.e., 
	$s_{\lfloor \frac{t}{T_s}-\frac{r_{\textrm{U},i}}{cT_s}\rfloor}\approx
	s_{\lfloor \frac{t}{T_s}-\frac{r_{\textrm{U}}}{cT_s}\rfloor}$,
	$w_i\left(t-\frac{r_{\textrm{U},i}}{c}\right)\approx w_i\left(t-\frac{r_{\textrm{U}}}{c}\right)$, and
	$\alpha(r_{\textrm{U},i})\approx \alpha(r_{\textrm{U}})$. Finally, ~\eqref{eq:yu4} is derived by expanding $r_{\textrm{U},ij}$ according to~\eqref{eq:ruij} and then ignoring the second-order small term $-j2\pi\frac{\Delta f_i\cdot \mathcal{L}_p(i,\theta)}{c}$.
	
	According to~\eqref{eq:FDA}, the term discarding the symbol index term $s_{\lfloor \frac{t}{T_s}-\frac{r_{\textrm{U}}}{cT_s}\rfloor}$ is the same as the array factor computed by previous FDA security work~\cite{cheng2019wfrft}. However, omitting the the $s_{\lfloor \frac{t}{T_s}-\frac{r_{\textrm{U}}}{cT_s}\rfloor}$ term is inappropriate it shows that the received symbol by Bob and Eve at time $t$ are different.
	Instead, we focus on one specific symbol transmitted Alice, and analyze how it is received by Bob and Eve. For a symbol $s_{n_0}$, Alice transmits the symbol at time $t={n_0}T_s$, while Bob and Eve receive this symbol at time $t={n_0}T_s+\frac{r_{\textrm{B}}}{c}$ and  $t={n_0}T_s+\frac{r_{\textrm{E}}}{c}$, respectively. Therefore, the corresponding received signals $y_\textrm{B}({n_0}T_s+\frac{r_{\textrm{B}}}{c},\theta,r_\textrm{B})$ and $y_\textrm{E}({n_0}T_s+\frac{r_{\textrm{E}}}{c},\theta,r_\textrm{E})$ can be expressed as
	\begin{subequations}
		\begin{align}
		\notag
		y_{\textrm{B}}\Big({n_0}T_s+\frac{r_{\textrm{B}}}{c},\theta&,r_\textrm{B}\Big)
		\\
		=s_{n_0}
		\alpha(r_{\textrm{B}})&
		\textrm{e}^{-j 2\pi f_0\left({n_0}T_s\right)}
		\sum_{i=1}^{N_t}w_i\left({n_0}T_s\right)
		\exp\left\{-j2\pi \left[f_0\frac{\mathcal{L}(i,\theta)}{c}- \Delta f_i\left({n_0}T_s\right) \right] \right\},
		\label{eq:n01}
		\\
		\notag y_{\textrm{E}}\Big({n_0}T_s+\frac{r_{\textrm{E}}}{c},\theta&,r_\textrm{E}\Big)\\
		=s_{n_0}
		\alpha(r_{\textrm{E}})&
		\textrm{e}^{-j 2\pi f_0\left({n_0}T_s\right)}
		\sum_{i=1}^{N_t}w_i\left({n_0}T_s\right)
		\exp\left\{-j2\pi \left[f_0\frac{\mathcal{L}(i,\theta)}{c}- \Delta f_i\left({n_0}T_s\right) \right] \right\}.
		\label{eq:n02}
		\end{align}
	\end{subequations}
	We observe that except for the distance-dependent path loss term $\alpha(r_{\textrm{U}})$ in~\eqref{eq:n01} and~\eqref{eq:n02}, the other terms are exactly the same. 
	Furthermore, since the precoding matrix satisfies $\sum_{i=1}^{N_t}|w_i|^2 = N_t$, the array factor term is upper-bounded by $N_t$, which is given by 
	\begin{equation}
	\left|\sum_{i=1}^{N_t}w_i\left({n_0}T_s\right)
	\exp\left\{-j2\pi \left[f_0\frac{\mathcal{L}(i,\theta)}{c}- \Delta f_i\left({n_0}T_s\right) \right] \right\}\right|^2\le \sum_{i=1}^{N_t}\left|w_i\left({n_0}T_s\right)
	\right|^2=N_t.
	\end{equation}
	Therefore, the maximum equivalent array gain of FDA is no difference from that of a conventional antenna array, i.e., by setting $\Delta f_i=0$.
	This leads to the conclusion that the FDA does not enhance the range security, which suggests the incorrect conclusion of previous studies on FDA~\cite{zheng2015multi,zheng2017physical,lv2018secure,ning2021joint}. 
	
	This conclusion can be further explained intuitively according to the principle of electromagnetic wave propagation. The carried symbols propagate at the speed of light, which has the same motion as the wavefront of the transmitted signal. Therefore, the beamforming gain corresponding to a symbol is determined by the beamforming pattern superimposed at the wavefront. 
	An important fact about wavefronts is that the frequency of EM waves only determines the speed of phase shift between consecutive wavefronts, but the phase of one wavefront does not change. Without loss of generality, we suppose the phase of the wavefront emitted at time $t$ is zero. Then, the wavefront phase emitted at time $t+\tau$ is given by $f\tau$. This phase of the wavefront does not change as it propagates outwards, which implies that changing the frequency of the antenna array, as FDA does, cannot affect the beamforming pattern superimposed by the wavefront. The signals superimposed at Bob can superimpose in the same pattern at Eve, and therefore it is useless to introduce a frequency offset among different antennas.
	
	From the perspective of communication system, it is worth noticing that a synchronizer is a fundamental part of the receiver in a communication system. Unlike time-invariant channels, the FDA system is not time-invariant, which makes it necessary to treat the synchronization problem carefully.
	Moreover, based on~\eqref{eq:FDA}, we discover that the addition of a frequency offset $\Delta f_i$ is equivalent with replacing $w_i(t)$ as $w_i(t)\cdot\exp\{-j2\pi \Delta f_i t\}$, i.e., adding the phase of the $i^{th}$ antenna by $\Delta f_i t$.
	This reminds us of the equivalence of adding a constant frequency and a time-varying phase from the perspective of an EM wave. Therefore, except for the different hardware implementation methods, a constant-frequency antenna array with time-variable phase shifters is equivalent to an FDA. Therefore, FDA cannot address the range security problem that the traditional phase-shifter-based multiple-antenna techniques cannot address.

	\section{Widely-Spaced Antenna Communications for Terahertz Range Security}\label{sec:wsa}
	Motivated by our conclusion that multiple antenna techniques operating in the far-field regions fail to provide THz range security in Sec.~\ref{sec:fda}, we discover multiple-antenna techniques operating in near-field regions.  
	In this section, we propose a THz WSA communication scheme to safeguard the THz range security.
	Specifically, We first develop the WSA model with hybrid beamforming and formulate a secrecy rate maximization problem. Then, we provide a hybrid beamforming design strategy for our WSA transmission scheme based on an NCOA algorithm. Finally, we analyze the trade-off between the enlarged antenna array size and the enhanced secrecy rate, which clarifies that the proposed scheme is realistic yet unique for THz band communications. 
	\begin{figure}[htbp]
		\centering
		\includegraphics[width=0.6\textwidth]{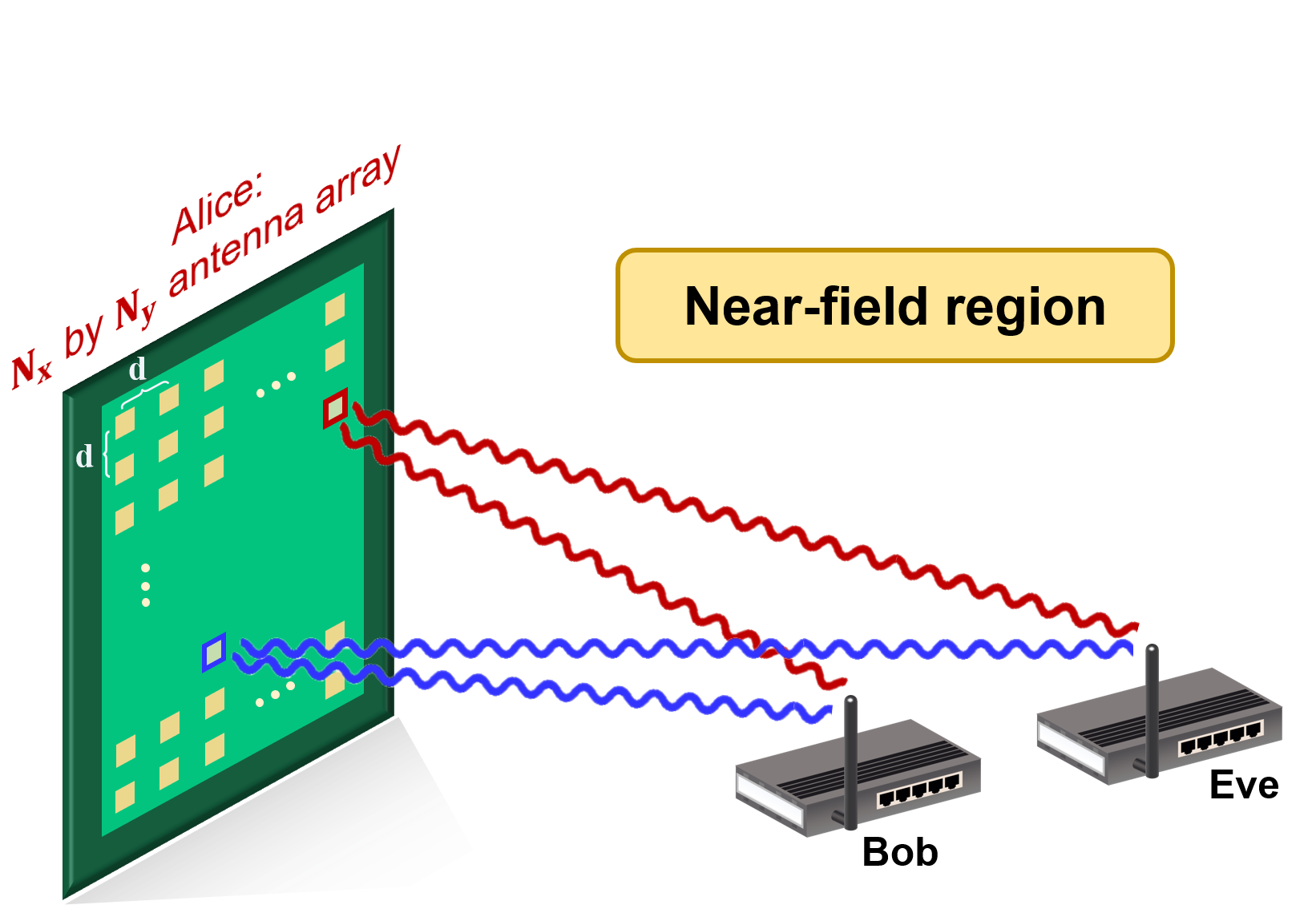}
		\caption{THz widely-spaced antenna array.}
		\captionsetup{font={footnotesize}}
		\label{fig:antenna-model}
	\end{figure}
	\subsection{Widely-Spaced Array and Hybrid Beamforming}
	The THz WSA model is illustrated in Fig.~\ref{fig:antenna-model}, where the spacing between antennas in the antenna array is widened such that the Alice-Bob distance and the Alice-Eve is comparable to the antenna size.
	The goal is that Bob and Eve can be located in the near-field region of the transmitter antenna array, and the closed-form expression for the residual propagation distance term in~\eqref{eq:rui_init} can be expressed as
	\begin{subequations}
		\begin{align}
		\mathcal{L}(i,\theta_{\textrm{U}},r_{\textrm{U}})
		&=r_{\textrm{U},i}-r_{\textrm{U}}\\
		&=\sqrt{\left(r_\textrm{U}\sin\theta_\textrm{U}-(i_{\textrm{row}}-1)d\right)^2+((i_{\textrm{col}}-1)d)^2+(r_\textrm{U}\cos\theta_\textrm{U})^2}-r_{\textrm{U}},
		\label{eq:SWP}
		\end{align}
	\end{subequations}
	where $i_{\textrm{row}}= (i-1) \mod N_y +1$ and $i_{\textrm{col}}=(i - i_{\textrm{row}})/ N_y +1$. 
	Remarkably, \eqref{eq:SWP} is the ground-truth model, since it considers the spherical-wave propagation nature for EM waves.
	
	For the beamformer design at the transmitter side, since traditional fully-digital beamformer requires the same number of radio frequency (RF) chains as antennas~\cite{ahmed2018survey}, which brings tremendous hardware cost for THz communications, hybrid beamforming technology is therefore proposed as a promising yet lower-complexity structure for UM-UPA~\cite{yan2020dynamic}. Specifically, the fully-connected hybrid beamforming structure is composed of a digital beamformer represented by $\mathbf{P}_{\textrm{D}}\in \mathbb{C}^{N_{\textrm{RF}}\times 1}$ and an analog beamformer expressed as $\mathbf{P}_{\textrm{A}}\in \mathbb{C}^{N_{t}\times N_{\textrm{RF}}}$. Each element of the matrices $\mathbf{P}_{\textrm{D}}$ and $\mathbf{P}_{\textrm{A}}$ represents the complex gain from the baseband signal to the RF chain, and from the RF chain to the antenna via phase shifters, respectively. $N_{\textrm{RF}}$ denotes the number of RF chains. The analog beamformer matrix is expressed in the form of
	\begin{equation}
	\mathbf{P}_{\textrm{A}}=\frac{1}{\sqrt{N_{t}}}
	\begin{pmatrix}
	\displaystyle
	\textrm{e}^{j\phi_{11}} & \cdots & \textrm{e}^{j\phi_{1N_{\textrm{RF}}}} \\
	\vdots & \ddots & \vdots \\
	\textrm{e}^{j\phi_{N_{t}1}} & \cdots & \textrm{e}^{j\phi_{N_{t}N_{\textrm{RF}}}}
	\end{pmatrix},
	\label{eq:PA}
	\end{equation}
	where $\phi_{pq}\in [0,2\pi), p\in\{1,\cdots,N_{t}\}, q\in\{1,\cdots, N_{\textrm{RF}}\}$ denotes the phase output of the phase shifter which connects the $p^{\textrm{th}}$ RF chain with the $q^{\textrm{th}}$ antenna. We assume that the hybrid beamforming does not affect the transmit power, i.e., $\| \mathbf{P}_{\textrm{A}} \mathbf{P}_{\textrm{D}} \|^2=1$.
	The overall beamforming vector is denoted by \begin{equation}
	\mathbf{w}=\mathbf{P}_{\textrm{A}} \mathbf{P}_{\textrm{D}} \triangleq[w_1,\cdots,w_{N_{t}}]^T.
	\end{equation}
	We assume that the carrier frequency of each antenna in the WSA is uniform and denoted as $f_0$.
	The received signal at the receiving node $U$ after down-converting and synchronization of symbol $n_0$ can be expressed as 
	\begin{equation}
	y_{\textrm{U}}(n_0,\theta_{\textrm{U}},r_{\textrm{U}})=\sqrt{\frac{P}{T_s}} s_{n_0} \alpha(r_{\textrm{U}})\sum_{i=1}^{N_t}w_i\exp\left\{j2\pi f_0 \frac{\mathcal{L}(i,\theta_{\textrm{U}},r_{\textrm{U}})}{c}\right\}+n(t),
	\end{equation}
	or equivalently we can express the received signal in matrix form as
	\begin{equation}
	y_{\textrm{U}}(n_0,\theta_{\textrm{U}},r_{\textrm{U}})=\sqrt{\frac{P}{T_s}} s_{n_0}\alpha(r_{\textrm{U}}) \mathbf{w}^T\mathbf{H}_{\textrm{U}}+n(t),
	\end{equation}
	where the channel combining matrix $\mathbf{H}_{\textrm{U}}$ is expressed as 
	\begin{equation}
	\mathbf{H}_{\textrm{U}}=\Big[\textrm{e}^{j2\pi\frac{\mathcal{L}(1,\theta_{\textrm{U}},r_{\textrm{U}})}{c}},\cdots,\textrm{e}^{j2\pi\frac{\mathcal{L}(N_t,\theta_{\textrm{U}},r_{\textrm{U}})}{c}}\Big]^T.
	\label{eq:Hr}
	\end{equation}
	
	\subsection{ Problem Formulation: Secrecy Rate Maximization}\label{sec:SRM}
	The secrecy rate, defined as the capacity difference between Bob's and Eve's channels, is used as the performance metric of our THz system model.
	As the focus of our THz secure communication system, the aim is to design transmit power $P$, hybrid beamformer, i.e., $\mathbf{P}_{\textrm{A}}$ and $\mathbf{P}_{\textrm{D}}$, to maximize the secrecy rate. Thus, we formulate the hybrid secure beamforming design problem as a secrecy rate maximization (SRM) optimization problem $\textrm{Q}_1$ as 
	\begin{subequations}
		\label{eq:secrecy rate max}
		\begin{flalign}
		\textbf{Q}_1: \max_{P,\mathbf{P}_{\textrm{D}}, \mathbf{P}_{\textrm{A}}}\quad & R_s=\left[ \log\left(1+\frac{P\left|\mathbf{w}^\dagger\mathbf{H}_{\textrm{B}}\right|^2 \alpha(r_{\textrm{B}})^2}{\sigma^2}\right)
		-\log\left(1+\frac{P\left|\mathbf{w}^\dagger\mathbf{H}_{\textrm{E}}\right|^2 \alpha(r_{\textrm{E}})^2}{\sigma^2}\right)\right]^+
		\label{eq:Rsmax:Rs}
		\\
		\text{s.t.}\quad 
		& P\le P_{\textrm{Tx}},
		\label{eq:Rsmax:P}
		\\
		& \| \mathbf{P}_{\textrm{A}} \mathbf{P}_{\textrm{D}} \|^2=1,
		\label{eq:Rsmax:PAPD}
		\\
		&\mathbf{P}_{\textrm{A}}=\frac{1}{\sqrt{N_{t}}}
		\begin{pmatrix}
		\displaystyle
		\textrm{e}^{j\phi_{11}} & \cdots & \textrm{e}^{j\phi_{1N_{\textrm{RF}}}} \\
		\vdots & \ddots & \vdots \\
		\textrm{e}^{j\phi_{N_{t}1}} & \cdots & \textrm{e}^{j\phi_{N_{t}N_{\textrm{RF}}}} 
		\end{pmatrix},
		\label{eq:Rsmax:PA}
		\end{flalign}
	\end{subequations}
	where \eqref{eq:Rsmax:Rs} denotes the secrecy rate of the THz secure communication system, and the operation $[x]^+\triangleq\max(x,0)$. \eqref{eq:Rsmax:P} states the transmit power constraint with the maximum power $P_{\textrm{Tx}}$. $\eqref{eq:Rsmax:PAPD}$ describes the normalization constraint of a hybrid beamformer. The maximal secrecy rate is denoted by $R_s^{\textrm{opt}}$ and the optimal transmit beamformer is represented by $\mathbf{w}^{\textrm{opt}}=[\mathbf{P}_{\textrm{A}}^{\textrm{opt}}\mathbf{P}_{\textrm{D}}^{\textrm{opt}}]$.
	Solving the problem $\textbf{Q}_1$ yields the optimal hybrid beamforming strategy achieving maximum secrecy. Due to the non-convexity of~\eqref{eq:Rsmax:Rs} with respect to $\mathbf{w}$, the optimization problem $\textbf{Q}_1$ is non-convex and cannot be tackled directly.
	Remarkably,	according to~\eqref{eq:Hr}, for the range security problem where $\theta_{\textrm{B}}=\theta_{\textrm{E}}$, if the far-field approximation $\mathcal{L}(i,\theta_{\textrm{U}},r_{\textrm{U}})\approx -(i_{\textrm{row}}-1)d\sin\theta_{\textrm{U}}$ is applied, we have $\mathcal{L}(i,\theta_{\textrm{B}},r_{\textrm{B}})=\mathcal{L}(i,\theta_{\textrm{E}},r_{\textrm{E}})$. In such case, Bob's and Eve's channels are perfectly correlated, i.e., $\mathbf{H}_{\textrm{B}}=\mathbf{H}_{\textrm{E}}$. As a result, their beamforming gains satisfy 
	$\left|\mathbf{w}^\dagger\mathbf{H}_{\textrm{B}}\right|^2=\left|\mathbf{w}^\dagger\mathbf{H}_{\textrm{E}}\right|^2$. Therefore, by considering that $\left|\mathbf{w}^\dagger\mathbf{H}_{\textrm{U}}\right|^2\in [0,N_{t}]$, the maximum secrecy rate is upper bounded by $\left[ \log\left(1+\frac{P_\textrm{Tx} N_{t} \alpha(r_{\textrm{B}})^2}{\sigma^2}\right)
	-\log\left(1+\frac{P_\textrm{Tx} N_{t} \alpha(r_{\textrm{E}})^2}{\sigma^2}\right)\right]^+$, which is exactly the maximum secrecy rate without adopting any security technique.

	\subsection{Hybrid Beamforming Design: NCOA Algorithm}
	We design a hybrid beamformer to maximize the secrecy rate in~\eqref{eq:Rsmax:Rs} for our WSA transmission scheme.
	Given the transmission distances $r_{\textrm{B}}$ and $r_{\textrm{E}}$, solving the SRM problem $\textbf{Q}_1$ yields our hybrid beamformer design. We propose an NCOA algorithm to solve this non-convex problem. First, we temporarily relax the non-convex constraint to transform $\textbf{Q}_1$ into a convex optimization problem $\textbf{Q}_2$. A globally optimal solution $\tilde{\mathbf{w}}$ to $\textbf{Q}_2$ is then derived, referred to as the non-constrained optimum. Finally, by reconsidering the non-convex constraint, an epsilon-convergent sub-optimal solution is achieved by approaching $\tilde{\mathbf{w}}$.  
	\subsubsection{Problem Transformation and Non-constrained Optimal Solution}
	We first temporarily neglect the constraint $\eqref{eq:Rsmax:PA}$. As a result, the remaining problem is converted to a convex optimization problem where a global optimum can be derived in a closed form. By maximizing the transmit power $P=P_\textrm{Tx}$ and allowing the condition that $\|w\|^2=1$, 
	the secrecy rate in \eqref{eq:Rsmax:Rs} is calculated as 
	\begin{equation}
	R_s=\log
	\left[
	1+\frac{\mathbf{w}^\dagger\left(\mathbf{H}_\textrm{B}\mathbf{H}_\textrm{B}^\dagger \cdot \alpha(r_\textrm{B})^2-\mathbf{H}_\textrm{E}\mathbf{H}_\textrm{E}^\dagger \cdot \alpha(r_\textrm{E})^2\right)\mathbf{w}}{\mathbf{w}^\dagger
		\left(\ffrac{\sigma^2}{P_{\textrm{Tx}}}\mathbf{I}+\mathbf{H}_\textrm{E}\mathbf{H}_\textrm{E}^\dagger \cdot \alpha(r_\textrm{E})^2\right)
		\mathbf{w}}
	\right],
	\end{equation}
	where $\mathbf{I}\in\mathbb{R}^{N_{t}\times N_{t}}$ represents a $N_{t}\times N_{t}$ identity matrix. Let $\textbf{A}\triangleq \mathbf{H}_\textrm{B}\mathbf{H}_\textrm{B}^\dagger \alpha(r_\textrm{B})^2-\mathbf{H}_\textrm{E}\mathbf{H}_\textrm{E}^\dagger \alpha(r_\textrm{E})^2$ and $\textbf{B}\triangleq \frac{\sigma^2}{P_{\textrm{Tx}}}\mathbf{I}+\mathbf{H}_\textrm{E}\mathbf{H}_\textrm{E}^\dagger \alpha(r_\textrm{E})^2$, maximizing $R_s$ is equivalent to maximizing the generalized Rayleigh quotient $\lambda_{\Sigma}\triangleq\frac{\mathbf{w}^\dagger \textbf{A}\mathbf{w}}{\mathbf{w}^\dagger \textbf{B}\mathbf{w}}$. 	  
	The transformation of the optimization problem from $\textbf{Q}_1$ to $\textbf{Q}_2$ is depicted in the following theorem. 
	
	\textbf{Theorem 1:}
	\it The SRM problem $\textbf{Q}_1$ without the non-convex constraint \eqref{eq:Rsmax:PAPD} is equivalent to the problem $\textbf{Q}_2$
	\begin{subequations}
		\label{eq:Q2}
		\begin{flalign}
		\textbf{Q}_2: \max_{\mathbf{w}}\quad & \lambda_{\Sigma}=|\langle\mathbf{w}',\mathbf{v}^{(a)}\rangle|^2\lambda_a+|\langle\mathbf{w}',\mathbf{v}^{(b)}\rangle|^2\lambda_b
		\label{eq:Q2:max}
		\\
		& \| \mathbf{w} \|^2=1,
		\label{eq:Q2:w}
		\end{flalign}
	\end{subequations}
	where $\mathbf{w}'=\frac{\textbf{B}^{1/2}\mathbf{w}}{\|\textbf{B}^{1/2}\mathbf{w}\|}$, $\lambda_a$ and $\lambda_b$ denote the only two non-zero eigenvalues of the matrix $\textbf{B}^{-\frac{1}{2}}\textbf{A}\textbf{B}^{-\frac{1}{2}}$ with $\lambda_a\ge \lambda_b$. $\mathbf{v}^{(a)}$ and $\mathbf{v}^{(b)}$ represent their corresponding normalized eigenvectors, respectively.
	
	Proof: 
	\rm The detailed proof is provided in Appendix \ref{ap2}.
	$\hfill\blacksquare$
	
	Since $|\langle\mathbf{w}',\mathbf{v}^{(a)}\rangle|\le |\langle\mathbf{v}^{(a)},\mathbf{v}^{(a)}\rangle|=1$, the non-constraint optimal solution to $\textbf{Q}_2$ is achieved when $\mathbf{w}'$ coincides with the eigenvector of the maximum eigenvalue $\mathbf{v}^{(a)}$. This leads the non-constrained optimal solution as
	\begin{equation}
	\tilde{\mathbf{w}}=\frac{\mathbf{B}^{-\frac{1}{2}}\mathbf{v}^{(a)}}{\|\mathbf{B}^{-\frac{1}{2}}\mathbf{v}^{(a)}\|},\quad \lambda_{\Sigma}=\lambda_{a},\quad R_s=\log_2(1+\lambda_{a}),
	\label{eq:glo_opt}
	\end{equation}
	where setting the beamforming matrix as $\hat{\mathbf{w}}$ achieves the maximum secrecy rate, which is an upper bound for the secrecy rate maximization problem $\textbf{Q}_1$. 
	\subsubsection{NCOA Algorithm}
	Given the non-constrained optimal solution $\tilde{\mathbf{w}}$ obtained in~\eqref{eq:glo_opt}, our aim is to approach $\mathbf{w}=[\mathbf{P}_{\textrm{A}}^{\textrm{opt}}\mathbf{P}_{\textrm{D}}^{\textrm{opt}}]$ to $\tilde{\mathbf{w}}$ under the analog beamforming constraint \eqref{eq:Rsmax:PA}. In~\cite{yan2020dynamic}, it is shown that when $N_{\textrm{RF}}\ge 2$, there exists a solution $\mathrm{P}_{\textrm{A}}$ and $\mathbf{P}_{\textrm{D}}$ that satisfies $[\mathbf{P}_{\textrm{A}} \mathbf{P}_{\textrm{D}} ]=\tilde{\mathbf{w}}$.
	Therefore, we divide the two cases, (i) the hybrid beamforming case where $N_{\textrm{RF}}\ge 2$, and (ii) the fully-analog (FA) beamforming case where $N_{\textrm{RF}}=1$, as follows.
	
	Case (i): For the hybrid beamforming case $N_{\textrm{RF}}\ge 2$, fortunately, it is possible to discover a feasible solution satisfying the constraint \eqref{eq:Rsmax:PA} and $\tilde{\mathbf{w}}=[\mathbf{P}_{\textrm{A}} \mathbf{P}_{\textrm{D}} ]$, which is expressed as
	\begin{subequations}
		\begin{align}
		\mathbf{P}_\textrm{A}^{\textrm{HB}}&=\left[\textrm{e}^{j\angle\tilde{\mathbf{w}}+\arccos\frac{\|\tilde{\mathbf{w}}\|}{2}}, \textrm{e}^{j\angle\tilde{\mathbf{w}}-\arccos\frac{\|\tilde{\mathbf{w}}\|}{2}},\cdots, 0\right]^T,\\\
		\mathbf{P}_\textrm{D}^{\textrm{HB}}&=\frac{1}{\sqrt{2}}\left[1,1,0,\cdots, 0\right]^T,
		\end{align}
		\label{eq:PAPDHB}
	\end{subequations}
	where $\angle(\cdot)$ returns the angle of the input complex element. 
	By setting the hybrid beamformer as \eqref{eq:PAPDHB}, the non-constrained optimal solution can be achieved, and the secrecy rate performance of the hybrid beamforming is maximized. 
	According to~\eqref{eq:PAPDHB}, it is worth noticing that by using the first two RF chains, the hybrid beamforming achieves the optimal performance, and using more than two RF chains cannot further improve the secrecy rate. This is consistent with an intuitive understanding that for one Tx and two single-antenna Bob and Eve with LoS transmissions, there are two SDoFs, which are equal to the DoF between Alice and Bob plus the DoF between Alice and Eve. Therefore, with two or more RF chains exceeding the number of SDoFs, the WSA communication system cannot further benefit from the increased RF chains. 
	
	Case (ii):
	For the fully-analog beamforming case with $N_{\textrm{RF}}=1$, since $\mathbf{P}_\textrm{D}^{\textrm{FA}}$ is a $1\times 1$ scalar number, we set $\mathbf{P}_\textrm{D}^{\textrm{FA}}=1$ without loss of generality. Then, the solution  $\mathbf{P}_\textrm{A}^{\textrm{FA}}=\tilde{\mathbf{w}}$ does not satisfy the constraint in~\eqref{eq:Rsmax:PA}. We use a gradient descent algorithm to recursively approach the optimal solution, where the gradient of the term \eqref{eq:Q2:max} versus $\mathbf{P}_\textrm{A}$ is represented by
	\begin{equation}
	\nabla\mathbf{P}_\textrm{A}^{\textrm{FA}}=\left[\frac{\partial \lambda_{\Sigma}}{\partial \phi_1}, \cdots, \frac{\partial \lambda_{\Sigma}}{\partial \phi_{N_{t}}}\right]^T,
	\label{eq:GD}
	\end{equation}
	where the term $\frac{\partial \lambda_{\Sigma}}{\partial \phi_i}$ is expressed as
	\begin{equation}
	\frac{\partial \lambda_{\Sigma}}{\partial \phi_i}=
	-\frac{2}{\sqrt{N_{t}}}\sum_{k\in\{a,b\}}\mathrm{Re}\left(\mathbf{v}_{k}^{\mathbf{B}\dagger}\mathbf{P}_\textrm{A}^{\textrm{FA}}\mathbf{e}^{-j\phi_i}\mathbf{e}_i\mathbf{v}_{k}^{\mathbf{B}}\right)\lambda_k,
	\label{eq:GDs}
	\end{equation}
	where $\mathbf{v}_{k}^{\mathbf{B}}=\frac{\textbf{B}^{-1/2}\mathbf{v}^{(k)}}{\|\textbf{B}^{-1/2}\mathbf{v}^{(k)}\|}$, and the vector $\mathbf{e}_i=[0,\cdots,0,1,0,\cdots,0]$ with the one in the $i^{\textrm{th}}$ column, $\mathrm{Re}(\cdot)$ returns the real part of a complex number. By recursively updating $\mathbf{P}_\textrm{A}^{\textrm{FA}}$ via computing the gradient in each step, a near-optimal solution for the fully-analog case can be achieved. 
	
	By combining the two solutions for the hybrid beamforming ($N_{\textrm{RF}}\ge 2$) and fully-analog beamforming ($N_{\textrm{RF}}=1$) cases, the NCOA algorithm for the hybrid beamforming design is summarized in \textbf{Algorithm 1}. $\epsilon$ denotes the step parameter controlling the convergence speed, and $\delta$ is the convergence threshold determining the convergence destination. In this work, we choose $\epsilon=10~\textrm{rad}$ and $\delta=0.003$, which results in good convergence performance. 
	By applying WSA communications for range security and considering the narrow beam nature with a UM-UPA, the range security for THz communications is thereby ensured. 
	\begin{algorithm}
		\caption{NCOA Algorithm} 
		\label{alg1}
		\hspace*{0.02in} {\bf Input:} 
		$r_{\textrm{B}}$, $r_{\textrm{E}}$, $\theta_{\textrm{B}}$, $\theta_{\textrm{E}}$ \\
		\hspace*{0.02in} {\bf Output:} 
		$P^{\textrm{opt}}$,
		$\mathbf{P}_\textrm{A}^{\textrm{opt}}$, $\mathbf{P}_\textrm{D}^{\textrm{opt}}$, $R_s^{\textrm{opt}}$
		\begin{algorithmic}[1]
			\State Compute the channel matrix $\mathbf{C}=\textbf{B}^{-\frac{1}{2}}\textbf{A}\textbf{B}^{-\frac{1}{2}}$ given $r_{\textrm{B}}$ and $r_{\textrm{E}}$;
			\State Perform eigenvalue decomposition (EVD) on $\mathbf{C}$ to compute $\lambda_a$, $\lambda_b$, $\mathbf{v}^{(a)}$, and $\mathbf{v}^{(b)}$;
			\State \textbf{Case I Hybrid beamforming:} when $N_{\textrm{RF}}\ge 2$,
			\State \quad Compute $\mathbf{P}_{\textrm{A}}^{\textrm{HB}}$,$\mathbf{P}_{\textrm{D}}^{\textrm{HB}}$ according to \eqref{eq:PAPDHB};
			\State \quad Compute $\lambda_\Sigma$ according to \eqref{eq:Q2:max};
			\State \quad Output
			$P^{\textrm{opt}}=P_{\textrm{Tx}}$,
			$\mathbf{P}_\textrm{A}^{\textrm{opt}}=\mathbf{P}_{\textrm{A}}^{\textrm{HB}}$, $\mathbf{P}_\textrm{D}^{\textrm{opt}}=\mathbf{P}_{\textrm{D}}^{\textrm{HB}}$, $R_s^{\textrm{opt}}=\log(1+\lambda_\Sigma)$;
			\State \textbf{Case II Fully-analog beamforming:} when $N_{\textrm{RF}}=1$,
			\State \quad $\mathbf{P}_\textrm{D}^{\textrm{FA}}=1$;
			\State \quad Randomly initialize $\mathbf{P}_\textrm{A,0}^{\textrm{FA}}$, and compute $\lambda_{\Sigma,0}$;
			\State \quad \textbf{repeat}
			\State \quad\quad Compute $\nabla\mathbf{P}_\textrm{A,i}^{\textrm{FA}}$ and $\lambda_{\Sigma,i}$ according to \eqref{eq:GD};
			\State \quad\quad Update $\mathbf{P}_\textrm{A,i+1}^{\textrm{FA}}\gets\mathbf{P}_\textrm{A,i}^{\textrm{FA}}+\epsilon \nabla\mathbf{P}_\textrm{A,i}^{\textrm{FA}}$;
			\State \quad\quad Compute $\lambda_{\Sigma,i+1}$;
			\State \quad\quad $i\gets i+1$;
			\State \quad \textbf{until} $\frac{|\lambda_{\Sigma,i+1}-\lambda_{\Sigma,i}|}{|\lambda_{\Sigma,i}|}<\delta$;
			\State \quad
			Output
			$P^{\textrm{opt}}=P_{\textrm{Tx}}$, $\mathbf{P}_\textrm{A}^{\textrm{opt}}=\mathbf{P}_{\textrm{A,i+1}}^{\textrm{FA}}$, $\mathbf{P}_\textrm{D}^{\textrm{opt}}=1$, $R_s^{\textrm{opt}}=\log(1+\lambda_{\Sigma,i+1})$;
		\end{algorithmic} 
	\end{algorithm}
	\subsection{Design Trade-off Between Array Size and Optimal Secrecy Rate}
	For a traditional antenna array where the antennas are not widely spaced, the far-field approximation is satisfied, which leads that the secrecy rate is upper-bounded by the secrecy capacity derived in~\eqref{eq:secrecy_rate_proof}. 
	Our WSA design widens the antenna spacing to ensure near-field communications. However, an ensuing drawback is that the array size becomes larger as the spacing between neighboring antennas increases, which makes the WSA design less attractive for practical applications.
	
	Thanks to the millimeter down to sub-millimeter THz wavelength, we clarify that THz communications can maintain a reasonable array size even when the antenna spacing is wide enough to enhance the secrecy rate. A sample comparison between mmWave and THz WSA design is provided as follows.
	For a $4\times 4$ micro-wave communication system where the wavelength is on the order of sub-meters, e.g., $0.3~\textrm{m}$ for the $1~\textrm{GHz}$ system, the near-field expression in~\eqref{eq:SWP} is valid when the side length of the array $3d\approx 4.0~\textrm{m}$, which is an impractically large antenna array. However, for THz communications, with a wavelength on the order of millimeter, e.g., $1~\textrm{mm}$ for a $300~\textrm{GHz}$ system, the near-field expression is valid when $3d\approx 0.08~\textrm{m}$. Therefore, a $0.08~\textrm{m}\times 0.08~\textrm{m}$ THz antenna array can well accommodate the near-field communication requirement.

	\section{Numerical Results}\label{sec:num}
	In this section, we evaluate the numerical results of the THz WSA scheme. First, we conduct Monte Carlo simulations on our recursion-based NCOA algorithm and demonstrate its convergence performance. Second, we analyze how system parameters, including the maximum transmit power, antenna spacing, and transmission distances affect the maximum secrecy rate. Moreover, the robustness of the proposed WSA scheme is studied, i.e., how the maximum secrecy rate degrades with the estimation error of the location of the Eve. Finally, the proposed scheme is compared with existing algorithms to demonstrate its improved range security performance. Unless specified, the system parameters used in the simulation are described in Table~\ref{Table}.
	\begin{table}[] 
		\caption{Simulation Parameters}
		\label{Table}
		\centering
		\begin{tabular}{p{2cm}p{7cm}p{2cm}p{2cm}} 
			\hline  
			\hline  
			\textbf{Notation} & \textbf{Definition} & \textbf{Value} & \textbf{Unit}\\  
			\hline 
			$P_\textrm{Tx}$ & Maximum transmit power & 10 & dBm\\
			$N_x$ & Number of row antennas & 32 & -\\
			
			$N_y$ & Number of column antennas & 32 & -\\ 
			
			$N_{t}$ & Total number of antennas & 1024 & -\\
			
			$\sigma^2$ & Noise variance & -80 & dBm\\
			
			$\theta$ & Propagation angle& $\pi/6$ & rad\\
			
			$f$ & Carrier frequency & 300 & GHz\\
			
			$r_{\textrm{B}}$ & Alice-Bob distance & 10 & m\\
			
			$r_{\textrm{E}}$ & Alice-Eve distance & 5 & m\\
			
			$\epsilon$ & Step parameter & 10 & rad\\
			
			$\delta$ & Convergence threshold & 0.003 & -\\
			\hline
			\hline  
		\end{tabular}  
	\end{table}
	\begin{figure*}
		\centering
		\subfigure[]{
			\includegraphics[width=0.44\textwidth]{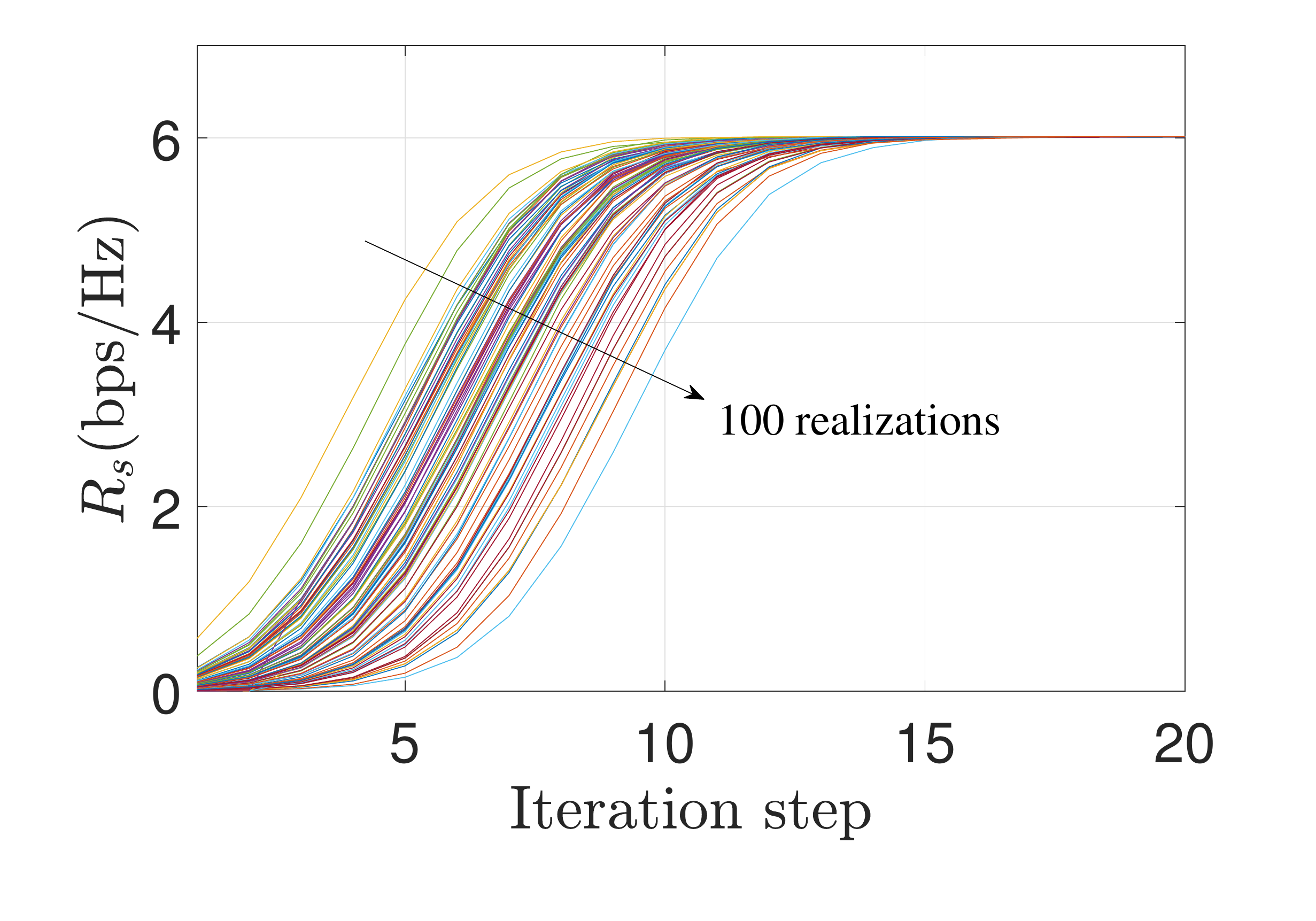}
			\label{fig:MonteCarlo_plot}
		}
		\subfigure[]{
			\includegraphics[width=0.46\textwidth]{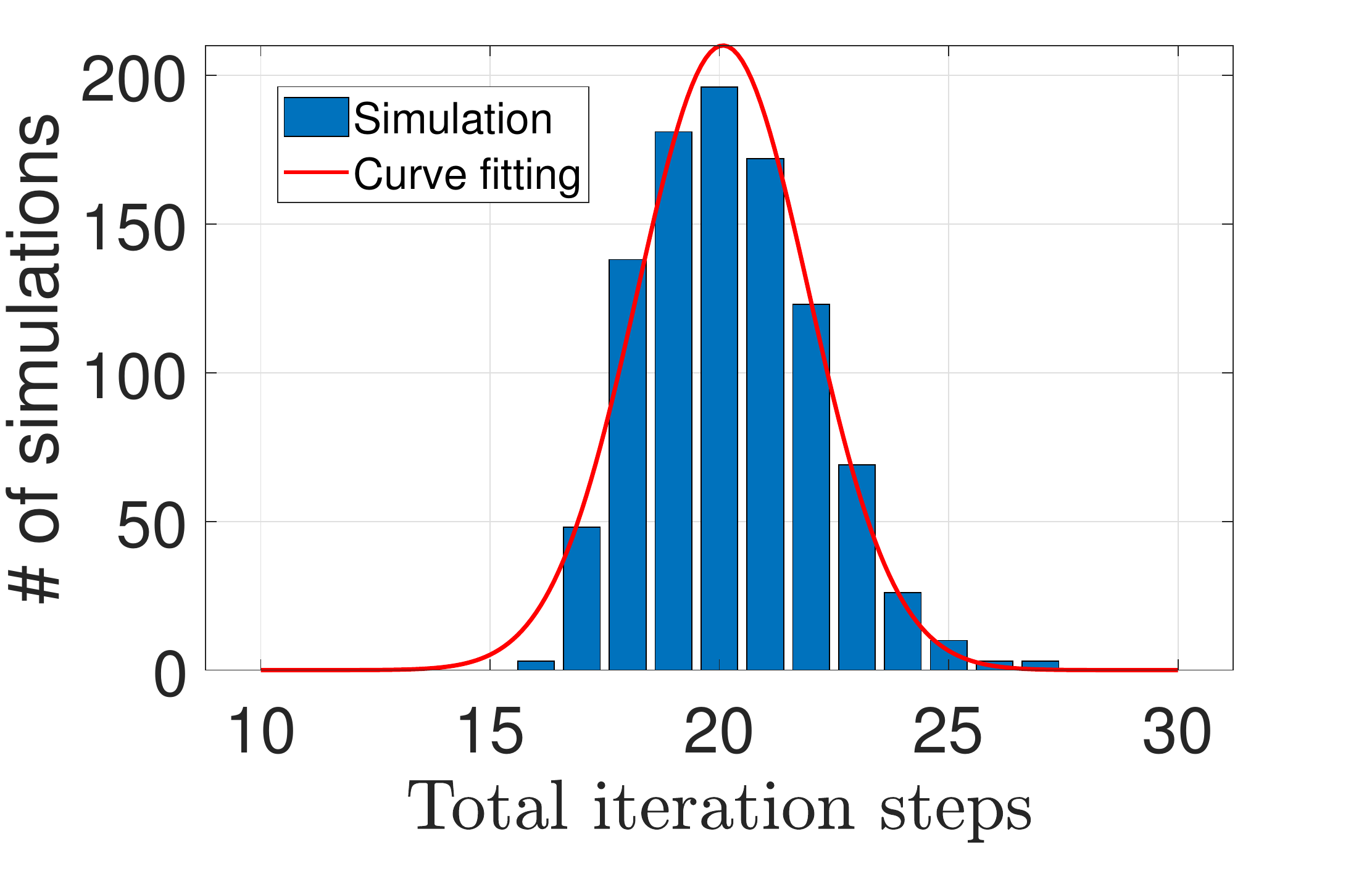}
			\label{fig:MonteCarlo_stat}
		}
		\caption{The Monte Carlo simulation on the convergence analysis of NCOA algorithm. (a) Convergence performance (100 simulations for illustration); (b) Statistical result of total iteration steps until convergence. }
		\label{fig:MonteCarlo}
		\captionsetup{font={footnotesize}}
	\end{figure*} 
	\subsection{Convergence Analysis of the NCOA Algorithm}
	To demonstrate the feasibility of the NCOA algorithm, we need to verify the convergence performance of the used gradient descent method. First, to verify the convergence, we perform Monte Carlo simulations on the initial value of $\mathbf{P}_{\textrm{A},0}^{\textrm{FA}}$ in step 9 of \textbf{Algorithm 1}. By setting different random initial values over 1,000 times, we plot the secrecy rates achieved by the solution versus the number of iterations, as depicted in Fig.~\ref{fig:MonteCarlo}. Fig.~\ref{fig:MonteCarlo_plot} plots 100 realizations randomly chosen from 1,000 realizations for illustration, where we observe that all simulations achieve nearly the same maximum secrecy rate within 20 iterations. This implies good convergence performance. 
	Fig.~\ref{fig:MonteCarlo_stat} presents statistical analysis on the distribution of the total number of iterations until convergence. Among 1,000 simulations, a Gaussian distribution with a mean of 20 is fitted, while no alias effect occurs.   
	\begin{figure*}[]
		\centering
		\includegraphics[width=0.45\textwidth]{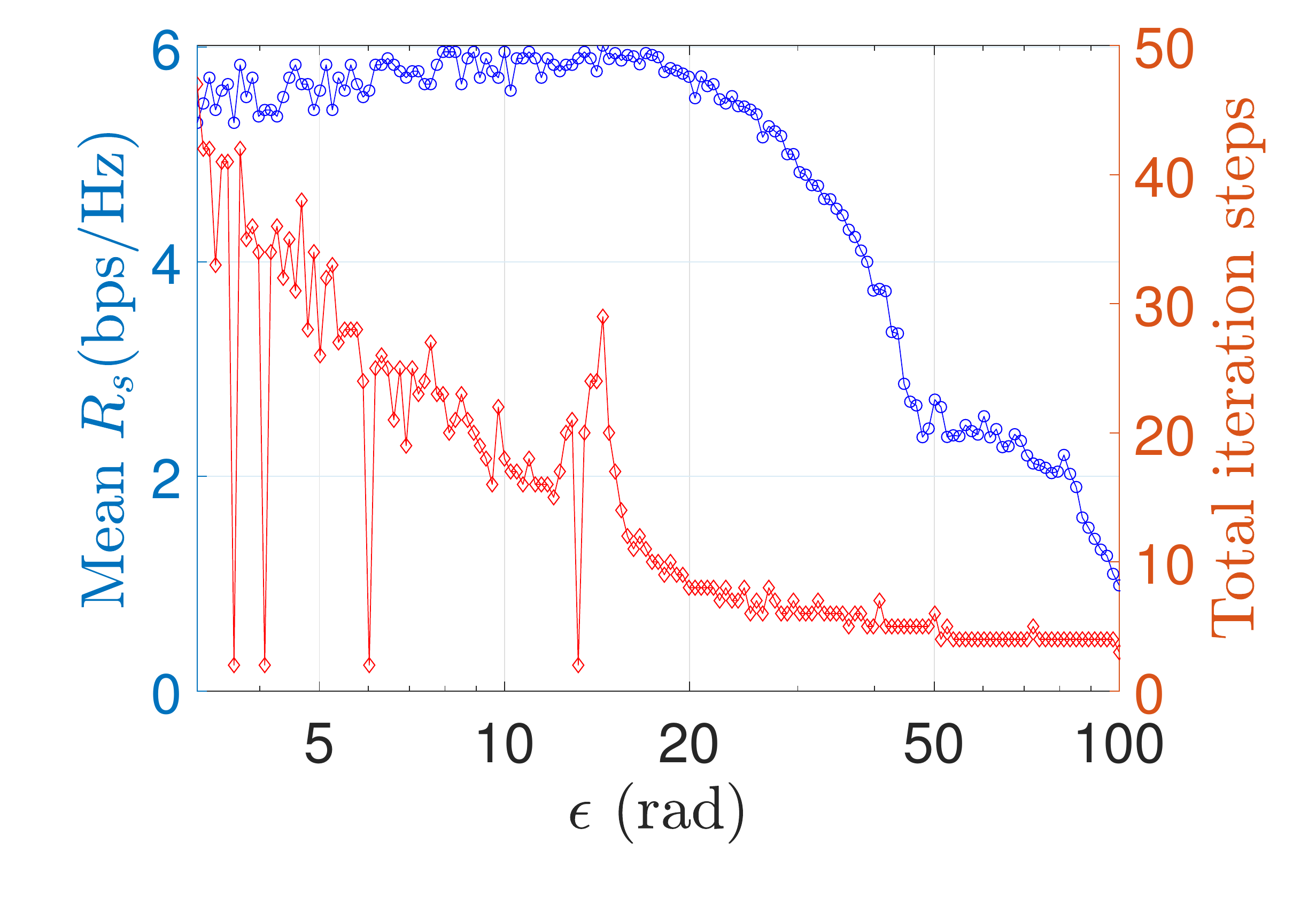}
		\caption{Mean secrecy rate and total iteration steps versus step parameter $\epsilon$.}
		\label{fig:Rsm_Nit}
		\captionsetup{font={footnotesize}}
	\end{figure*}
	
	In Fig.~\ref{fig:Rsm_Nit}, we explore how the step parameter $\epsilon$ in step 12 in Algorithm~1 affects the maximum secrecy rate and the total iteration step reaching the termination condition. For each $\epsilon$, 100 Monte Carlo simulations are performed to calculate the mean secrecy rate. We can observe that when $\epsilon<20$, the mean secrecy rate approximately reaches the maximum value. By contrast, when $\epsilon>20$, the mean secrecy rate decays rapidly. This is due to the fact that a large iteration step $\epsilon$ leads to an early convergence before the secrecy rate approaches the maximum. Moreover, the number of total iteration steps decreases as $\epsilon$ increases, as the right y-axis of Fig.~\ref{fig:Rsm_Nit} presents. The numerical results demonstrate that setting $\epsilon$ within [10, 20] results in a good balance between performance and complexity.
	\begin{figure*}
		\centering
		\subfigure[]{
			\label{fig:Rs_d_PTx} 
			\includegraphics[width=0.31\textwidth]{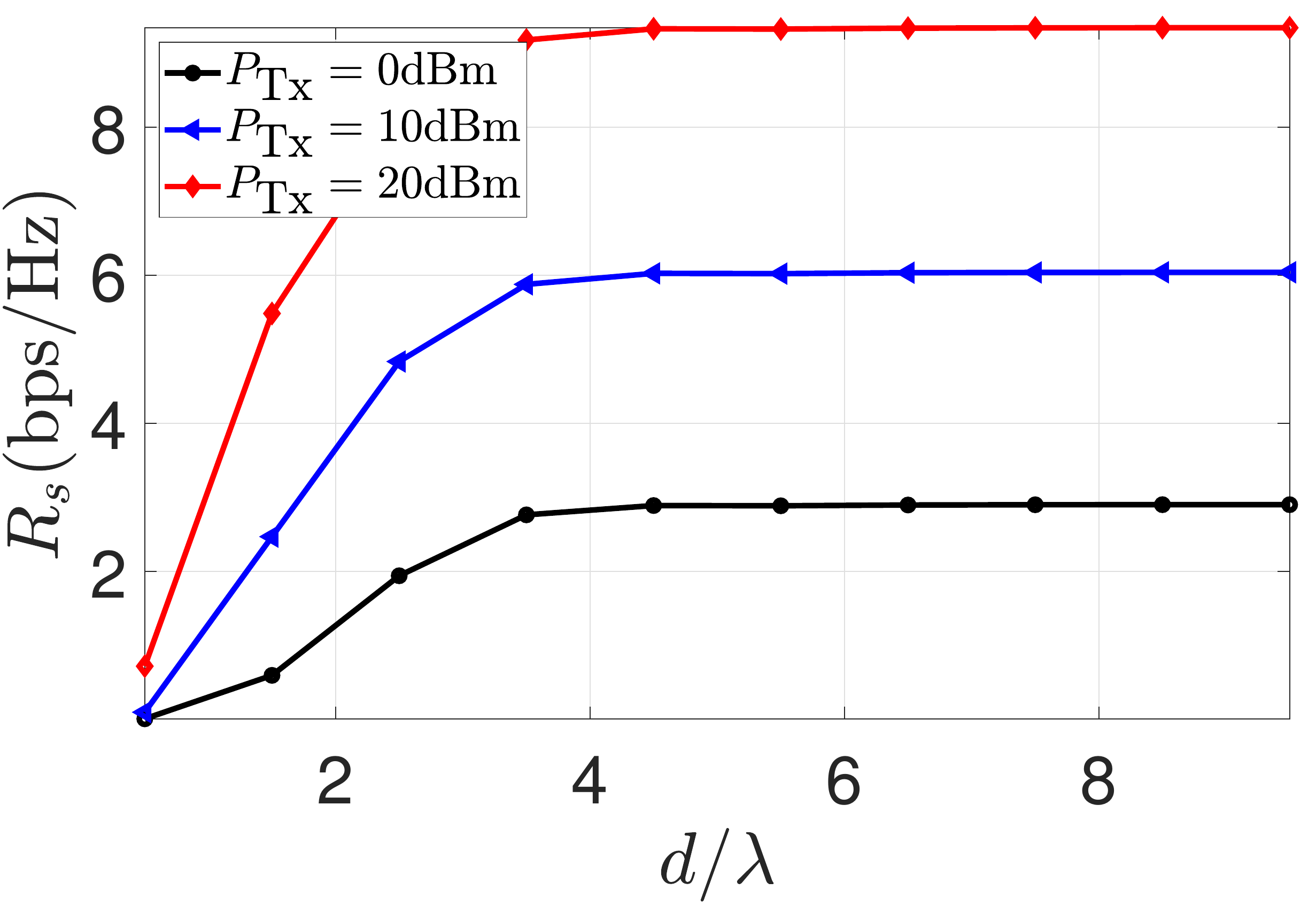}}
		\subfigure[]{
			\label{fig:Rs_DED_PTx} 
			\includegraphics[width=0.31\textwidth]{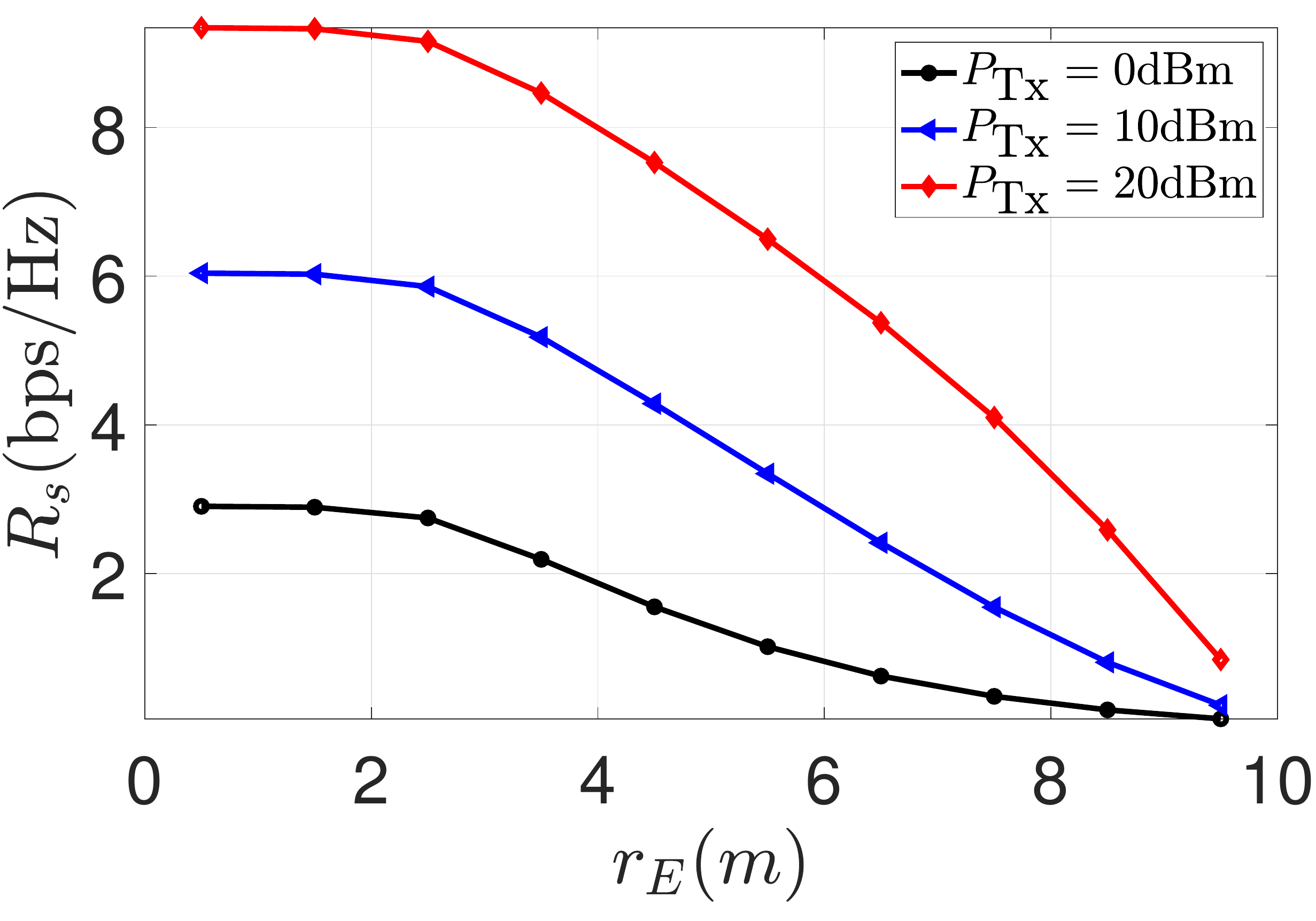}}
		\subfigure[]{
			\label{fig:Rs_Nt_PTx} 
			\includegraphics[width=0.31\textwidth]{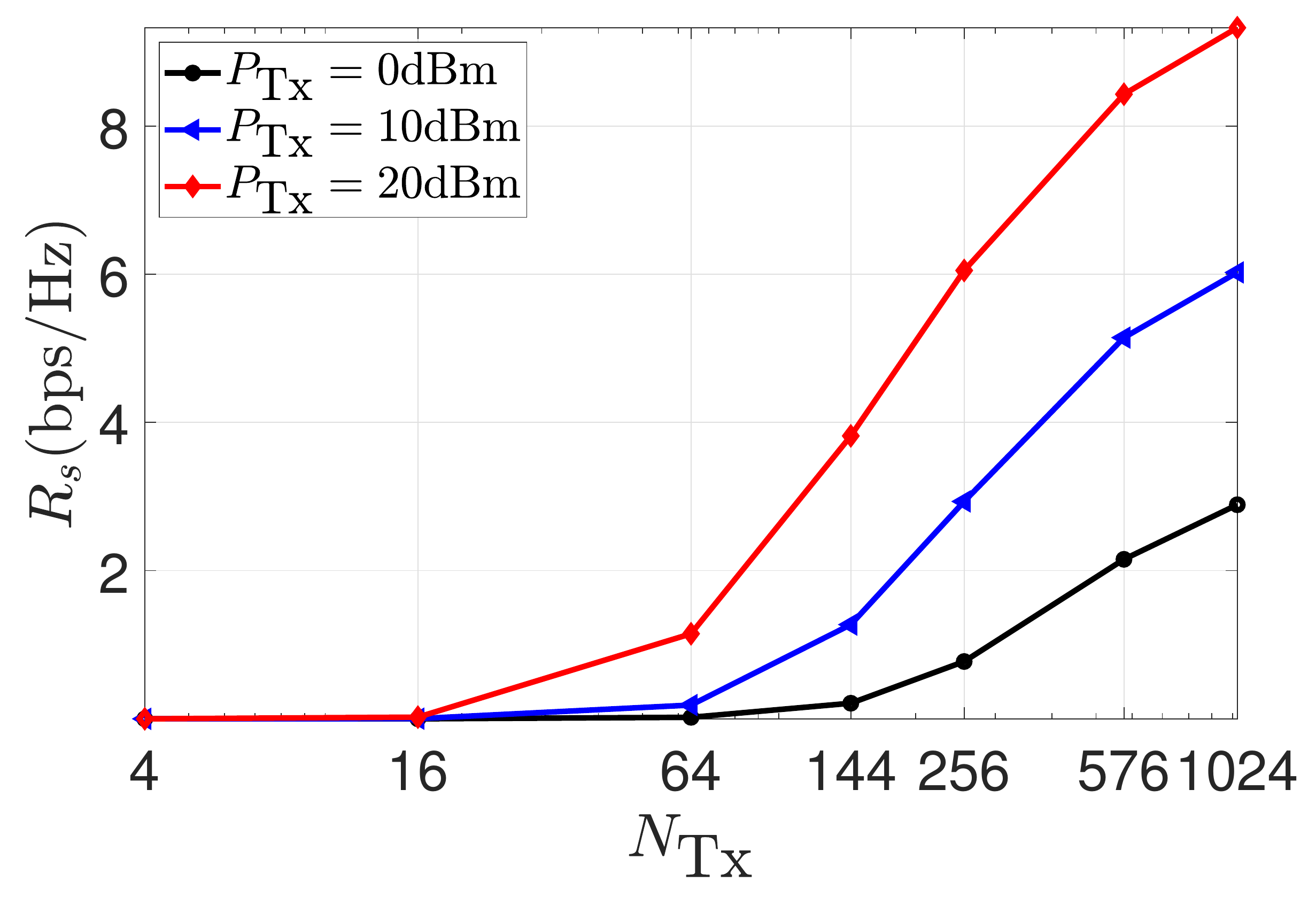}}
		\caption{Secrecy rate versus different system parameters ($N_{\textrm{RF}}=2$). (a) With varying antenna spacing $d$. (b) With varying Alice-Eve distance $r_{\textrm{E}}$. (c) With varying array size $N_t$. }
		\captionsetup{font={footnotesize}}
	\end{figure*}
	
	\begin{figure*}
		\centering
		\subfigure[]{
			\label{fig:robustness_DED} 
			\includegraphics[width=0.45\textwidth]{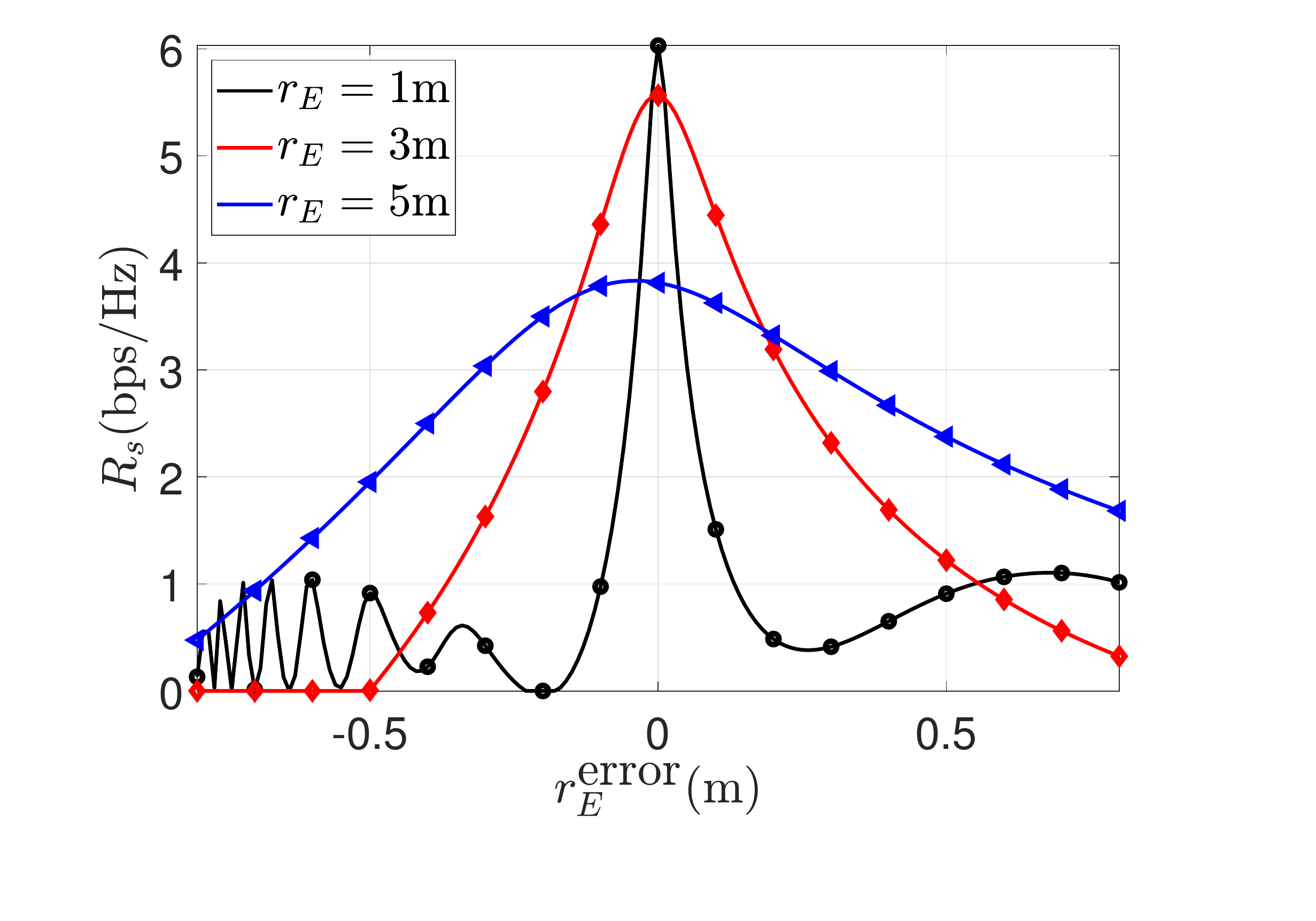}}
		\subfigure[]{
			\label{fig:robustness_theta} 
			\includegraphics[width=0.45\textwidth]{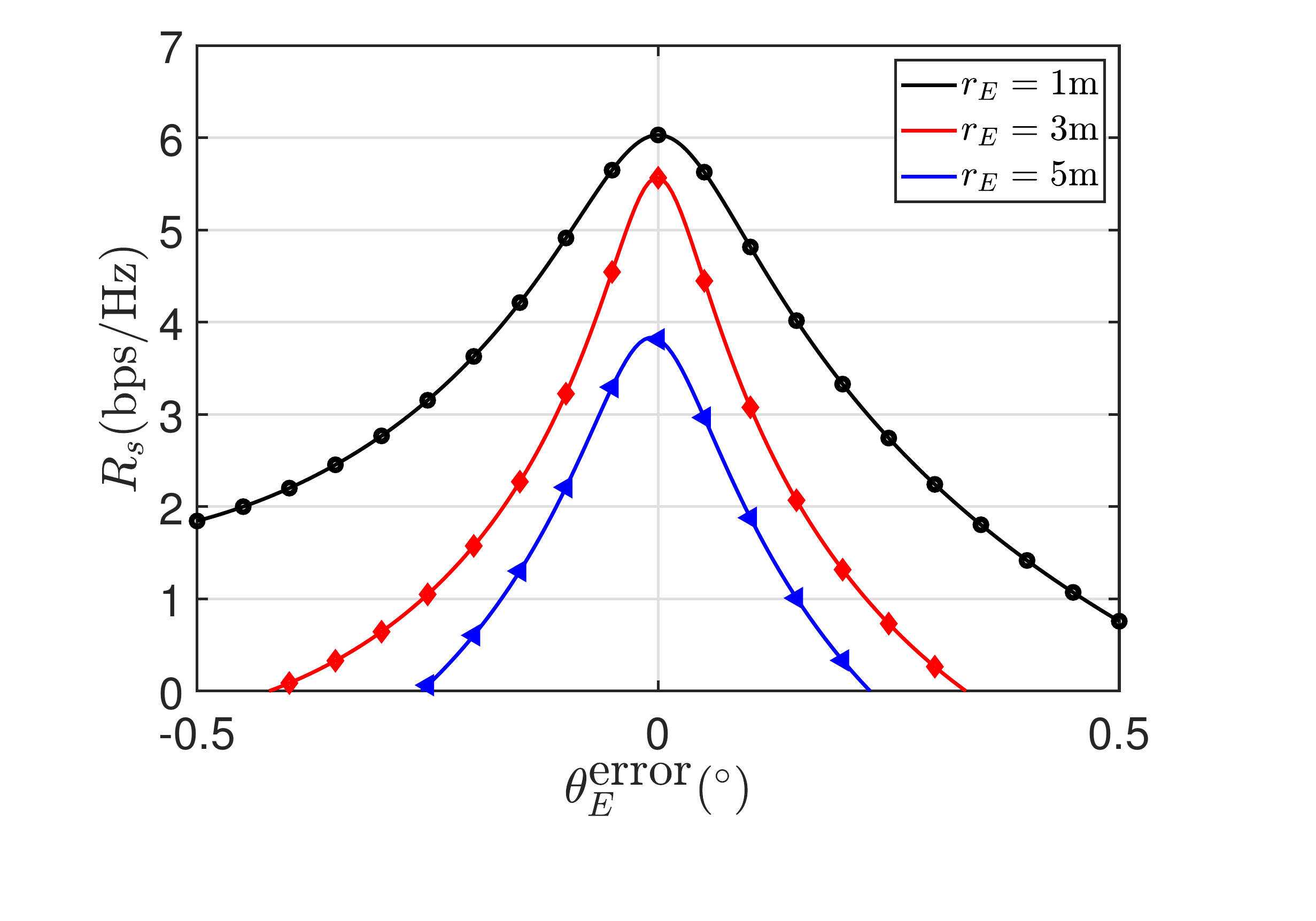}}
		\caption{Robustness of secrecy rate to the location estimation error of Eve's position. (a) Secrecy rate versus the distance estimation error $r_\textrm{E}^{\textrm{error}}$. (b)  Secrecy rate versus the angle estimation error $\theta_\textrm{E}^{\textrm{error}}$.}
		\captionsetup{font={footnotesize}}
		\label{fig:robustness}
	\end{figure*}
	\subsection{Maximum Secrecy Rate}
	We compute the maximum secrecy rate versus different system parameters for our proposed WSA scheme for THz range security. The maximum secrecy rates with different maximum transmit power values are shown in Fig.~\ref{fig:Rs_d_PTx} for the different antenna spacing, in Fig.~\ref{fig:Rs_DED_PTx} for the different eavesdropping distances, and in Fig.~\ref{fig:Rs_Nt_PTx} for the different antenna array sizes. In general, the secrecy rate increases with a wider antenna spacing, a nearer Eve distance, and larger array size. Moreover, with $d=5\lambda$, $r_{\textrm{E}}=5~\textrm{m}$, and $N_{t}=1024$, the secrecy rate with maximum transmit power of $10~\textrm{dBm}$ is $6~\textrm{bps/Hz}$. 
	An interesting phenomenon captured in Fig.~\ref{fig:Rs_DED_PTx} is that as $r_{\textrm{E}}$ becomes smaller, i.e., Eve approaches the Tx, the secrecy shows an increasing trend. This is explained that when Eve is nearer to Alice or, equivalently, farther from the B, Eve's channel is more uncorrelated with Bob's channel, which enhances the secrecy rate though Eve's channel gain increases. 
	\subsection{Robustness to Location Estimation Error of Eavesdroppers}
	The proposed NCOA algorithm needs to acquire Eve's location, i.e., the Alice-Eve distance and angle, and we estimate these two values as the input of our beamformer design.
	Therefore, We need to study the robustness of our scheme to the location estimation error of Eve, i.e., the robustness of the Alice-Eve distance and angle. The estimation error is defined as the difference between the estimated value and the nominal value, and the estimation errors of the Alice-Eve distance and angle are denoted as $r_{\textrm{E}}^{\textrm{error}}$ and $\theta_{\textrm{E}}^{\textrm{error}}$, respectively.
	The maximum secrecy rates versus $r_{\textrm{E}}^{\textrm{error}}$ and $\theta_{\textrm{E}}^{\textrm{error}}$ are plotted in Fig.~\ref{fig:robustness}.
	In our simulations, $r_\textrm{E}$ denotes the true Alice-Eve distance. In Fig.~\ref{fig:robustness_DED}, we observe that
	the secrecy rate degrades dramatically as the Alice-Eve distance estimation error increases. Moreover, as $r_{\textrm{E}}$ decreases, the degradation becomes severer with the estimation error at short transmission distances. This can be explained that at short distances, a small distance estimation error leads to a large phase error on each antenna.
	In Fig.~\ref{fig:robustness_theta}, we observe that for the different Alice-Eve distances, the maximum secrecy rate degrades by 10\% when the Alice-Eve angle has a $\pm 0.03^\circ$ error. 
	This result implies that the robustness of the WSA communication scheme to the Alice-Eve estimation error is highly dependent on the Alice-Eve distance. Moreover, it is suggested that the estimation error of the Alice-Eve angle should be less than $0.03^\circ$ to ensure a less than 10\% secrecy rate degradation.
	
	\subsection{Performance Comparison With Different Methods}
	\begin{figure*}
		\centering
		\subfigure[]{
			\includegraphics[width=0.45\textwidth]{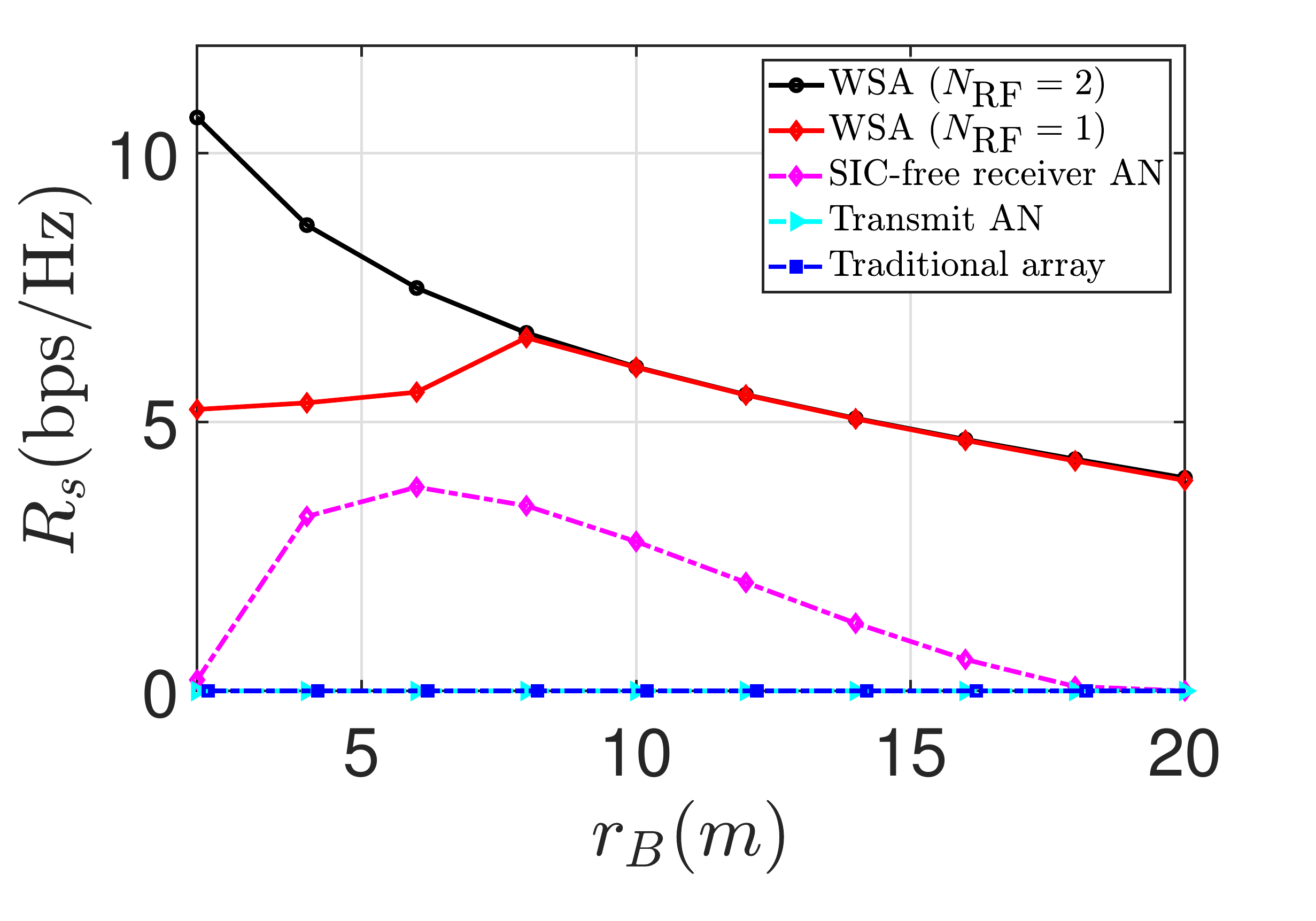}
			\label{fig:Rs_DLU}
		}
		\subfigure[]{
			\includegraphics[width=0.45\textwidth]{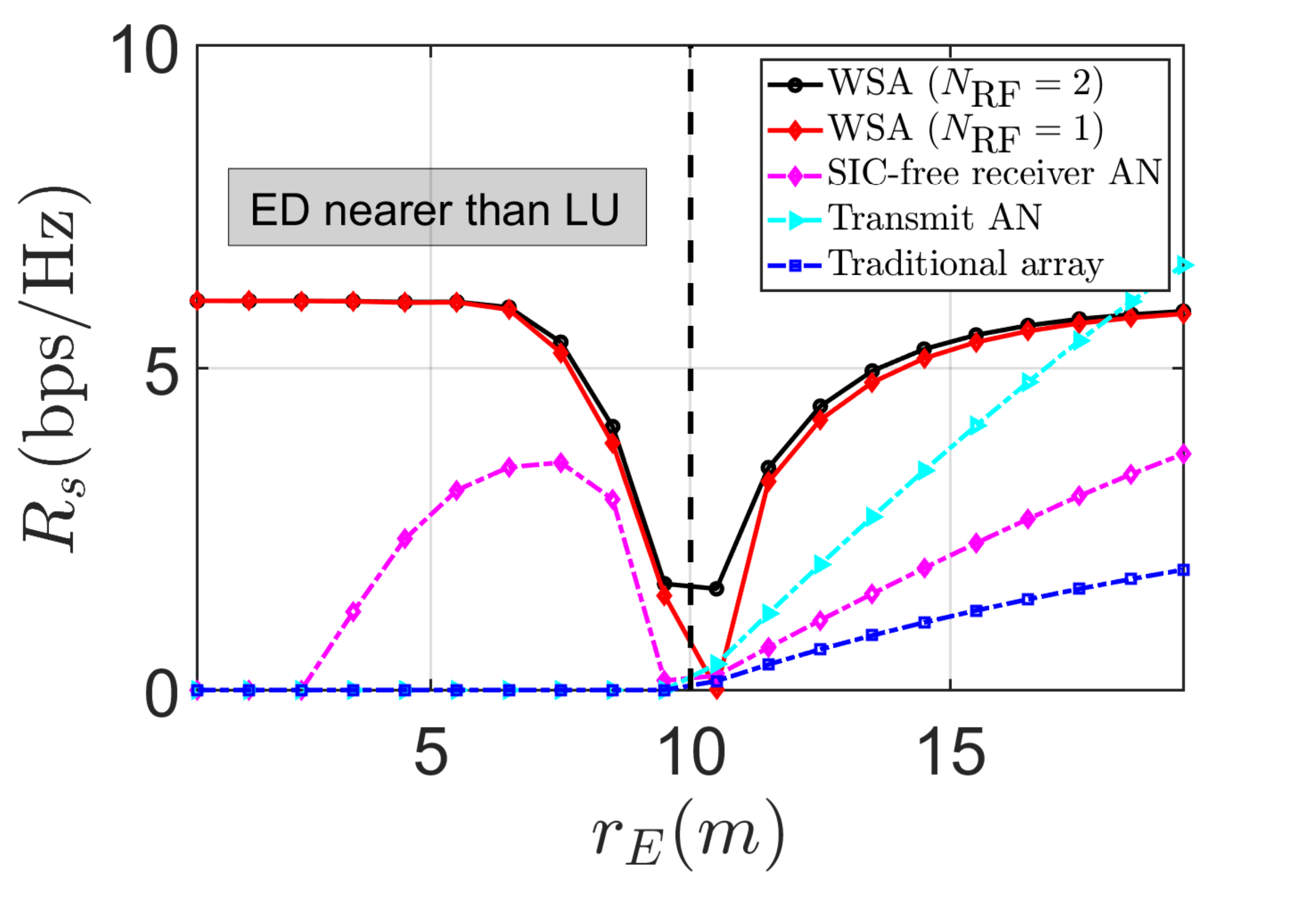}
			\label{fig:Rs_DED}
		}
		\caption{Performance comparison between different methods. (a) With varying Alice-Bob distance $r_\textrm{B}$. (b) With varying Alice-Eve distance $r_\textrm{E}$.}
		\captionsetup{font={footnotesize}}
		\label{fig:comparison}
	\end{figure*}
	We evaluate and compare the maximum secrecy rate achieved by our proposed WSA scheme with existing schemes in the literature. Specifically, the transmit AN scheme~\cite{lv2018secure}, the SIC-free receiver AN scheme~\cite{myarticle2}, and traditional antenna array hybrid beamforming method~\cite{yan2020dynamic} are compared in Fig.~\ref{fig:comparison}. The transmit power of the receiver AN signal for the SIC-free receiver AN scheme is 10~$\textrm{dBm}$.
	In Fig.~\ref{fig:Rs_DLU}, the secrecy rate versus the Alice-Bob distance is plotted.
	Moreover, the hybrid beamforming ($N_{\textrm{RF}}\ge2$) and fully-analog beamforming ($N_{\textrm{RF}}=1$) cases of our WSA secure beamforming scheme are computed to demonstrate the sub-optimal performance of fully-analog beamforming structure for the optimal hybrid beamforming case. 
	For the $N_{\textrm{RF}}\ge2$ case, the maximum secrecy rate decreases as $r_{\textrm{B}}$ increases. At $r_{\textrm{B}}=10~\textrm{m}$, the secrecy rate achieves $6~\textrm{bps/Hz}$. The secrecy rate of the $N_{\textrm{RF}}\ge1$ case, as an inferior but low-complexity architecture of the $N_{\textrm{RF}}=2$ case, can achieve near-optimal performance when $r_{\textrm{B}}$ is larger than $8~\textrm{m}$. This implies that the performance degradation due to the sub-optimality of hybrid beamforming becomes severer when the distance between Alice and Bob is small. Moreover, compared with the SIC-free receiver AN scheme~\cite{myarticle2}, the proposed WSA scheme can double the secrecy rate at $r_{\textrm{B}}=10~\textrm{m}$. This significant improvement originates from the different security-enhancing strategies used by the two schemes. Specifically, the WSA scheme directly mitigates the received power by Eve, while the SIC-free receiver AN is to jam Eve's signal. We observe that the secrecy rates achieved by the transmit AN scheme and traditional hybrid beamforming are nearly zero, which is consistent with our analysis in Sec.~\ref{sec:SRM} that the secure communication in the far field fails to achieve a positive secrecy rate.
	
	In Fig.~\ref{fig:Rs_DED}, we plot the maximum secrecy rate versus the different Alice-Eve distances varying from $0.5~\textrm{m}$ to $20~\textrm{m}$, where the Alice-Bob distance is $10~\textrm{m}$. We see that when $r_\textrm{E}$ increases from $0.5~\textrm{m}$ to $10~\textrm{m}$, the secrecy rate of WSA scheme decreases as $r_\textrm{E}$ increases. Moreover, the secrecy rate increases as $r_\textrm{E}$ increases when $r_\textrm{E}$ is larger than $10~\textrm{m}$. This is due to the fact that as $r_\textrm{E}$ approaches 10~\textrm{m}, Bob's channel and the Eve's channel become correlated, which leads that the WSA scheme cannot distinguish the two receivers in the range domain. 
	Moreover, the WSA scheme with two RF chains can achieve the secrecy rate of approximately $6~\textrm{bps/Hz}$, as $r_\textrm{E}$ approaches zero. However, this does not imply that the proposed scheme can effectively combat the extremely near eavesdropping since the robustness against the Alice-Eve distance estimation error degrades significantly as $r_\textrm{E}$ approaches zero.

	\section{Conclusion}\label{sec:concl}
	This paper investigates the multiple-antenna technologies for THz range security, including the FDA and WSA. Specifically, we first present a multiple-antenna-assisted THz range security model and theoretically derive the secrecy capacity of multiple antenna channels. We prove that the secrecy capacity for Bob and Eve in the far-field region is equal to the secrecy rate achieved by traditional beamforming schemes. Based on this conclusion, we revisited and revised the FDA range security model. Then, we proposed a WSA scheme and a hybrid beamforming design to enhance the secrecy rate. The WSA transmission is realized by increasing the antenna spacing to place Bob and Eve in the near-field region of the antenna array. For the optimal hybrid beamformer design, we develop an NCOA algorithm to achieve the closed-form optimal solution for the hybrid beamforming case and an epsilon-convergent sub-optimal solution for the fully-analog beamforming case, respectively. Numerical results verify the outstanding convergence of the NCOA algorithm and demonstrate that under the range security condition, the secrecy rate of the WSA communication scheme reaches $6~\textrm{bps/Hz}$ with $10~\textrm{dBm}$ transmit power. 
	\begin{appendices}
		\section{Proof of Theorem 1}\label{ap2}
		Due to the monotonicity of the logarithmic function, Maximizing $R_s$ in~\eqref{eq:Rsmax:Rs} is equivalent to maximizing $\frac{\mathbf{w}^\dagger \textbf{A}\mathbf{w}}{\mathbf{w}^\dagger \textbf{B}\mathbf{w}}$.
		First, we prove the existence of the matrix $\textbf{B}^{-\frac{1}{2}}$. Since for any non-zero vector $\textbf{x}$, we have $\textbf{x}^\dagger \textbf{B}\textbf{x}=\textbf{x}^\dagger 	\frac{\sigma^2}{P_{\textrm{Tx}}}\mathbf{I}\textbf{x}+\textbf{x}^\dagger\mathbf{H}_\textrm{E}\mathbf{H}_\textrm{E}^\dagger\textbf{x}\cdot \alpha(r_\textrm{E})^2=\frac{\sigma^2}{P_{\textrm{Tx}}}\|\textbf{x}\|^2+\|\mathbf{H}_\textrm{E}^\dagger\textbf{x}\|^2 \cdot \alpha(r_\textrm{E})^2> 0$, $\textbf{B}$ is positive definite, and $\textbf{B}^{-\frac{1}{2}}$ exists. Thus, we can assume $\mathbf{w}'=\frac{\textbf{B}^{1/2}\mathbf{w}}{\|\textbf{B}^{1/2}\mathbf{w}\|}$ and $\textbf{C}= \textbf{B}^{-\frac{1}{2}}\textbf{A}\textbf{B}^{-\frac{1}{2}}$, we have $\frac{\mathbf{w}^\dagger \textbf{A}\mathbf{w}}{\mathbf{w}^\dagger \textbf{B}\mathbf{w}}=\mathbf{w}'^\dagger\textbf{C}\mathbf{w}'$. 
		We first prove that $\textrm{rank}(\textbf{C})=2$, which implies that $\textbf{C}$ only has two non-zero eigenvalues.	  
		First, since the ranks of vector $\mathbf{H}_{\textrm{B}}$ and $\mathbf{H}_{\textrm{E}}$ are 1,   $\mathrm{rank}(\mathbf{H}_\textrm{B}\mathbf{H}_\textrm{B}^\dagger)=\mathrm{rank}(\mathbf{H}_\textrm{E}\mathbf{H}_\textrm{E}^\dagger)=1$. Therefore, $\mathrm{rank}(\textbf{A})=\mathrm{rank}(\mathbf{H}_\textrm{B}\mathbf{H}_\textrm{B}^\dagger \alpha(r_\textrm{B})^2-\mathbf{H}_\textrm{E}\mathbf{H}_\textrm{E}^\dagger \alpha(r_\textrm{E})^2)=2$. Second, since the identity matrix $\mathbf{I}$ has full rank and $\mathbf{H}_\textrm{E}\mathbf{H}_\textrm{E}^\dagger \alpha(r_\textrm{E})^2$ is a Hermitian matrix, the matrix $\textbf{B}$ as well as $\textbf{B}^{-1/2}$ have full rank. As a result, $\textbf{C}$ is a rank-2 matrix, and there are two non-zero eigenvalues. We denote the two eigenvalues as $\lambda_a$ and $\lambda_b$ with their corresponding eigenvectors $\mathbf{v}^{(a)}$ and $\mathbf{v}^{(b)}$.\
		
		Next, we decompose the vector $\mathbf{w}'$ onto the orthonormal basis, composed of all normalized eigenvectors of $\textbf{C}$ denoted by $\{\mathbf{v}^{(i)}\}, i=1,\cdots, N_{t}$. As
		$\mathbf{w}'=\sum_{i=1}^{N_{t}}\langle \mathbf{w}', \mathbf{v}^{(i)}\rangle \mathbf{v}^{(i)}$, we can write
		\begin{equation}
		\begin{aligned}
		\mathbf{w}'^\dagger \mathbf{C}\mathbf{w}'&=\left(\sum_{i=1}^{N_{t}}\langle \mathbf{w}', \mathbf{v}^{(i)}\rangle \mathbf{v}^{(i)}\right)\left(\mathbf{C}\sum_{j=1}^{N_{t}}\langle \mathbf{w}', \mathbf{v}^{(j)}\rangle \mathbf{v}^{(j)}\right)
		\\
		&=
		\left(\sum_{i=1}^{N_{t}}\langle \mathbf{w}', \mathbf{v}^{(i)}\rangle \mathbf{v}^{(i)}\right)\left(\sum_{j=1}^{N_{t}}\langle \mathbf{w}', \mathbf{v}^{(j)}\rangle \lambda_j\mathbf{v}^{(j)}\right)\\
		&=\sum_{i,j}\Big|\langle \mathbf{w}', \mathbf{v}^{(i)}\rangle\langle \mathbf{w}', \mathbf{v}^{(j)}\rangle\Big|\langle\mathbf{v}^{(i)}, \mathbf{v}^{(j)}\rangle\lambda_j\\
		&=\sum_{i=1}^{N_{t}}\Big|\langle \mathbf{w}', \mathbf{v}^{(i)}\rangle^2\Big|\lambda_i\\
		&=|\langle\mathbf{w}',\mathbf{v}^{(a)}\rangle|^2\lambda_a+|\langle\mathbf{w}',\mathbf{v}^{(b)}\rangle|^2\lambda_b,
		\end{aligned}
		\end{equation}
		which completes the proof.
		$\hfill\blacksquare$
		\rm

	\end{appendices}
	\bibliographystyle{IEEEtran}
	\bibliography{main}
\end{document}